\documentclass[aps,prapplied,twocolumn,groupedaddress,floatfix]{revtex4-2}

\usepackage{graphicx,epstopdf,siunitx,braket,xspace,amssymb,dsfont,bm,amsmath,bbold,xstring,placeins,xfrac}
\usepackage[utf8]{inputenc}

\newcommand{\mc}[1]{\ensuremath{\mathcal{#1}}}
\newcommand{\mr}[1]{\ensuremath{\mathrm{#1}}}
\newcommand{\eps}[0]{\ensuremath{\epsilon}\xspace}
\newcommand{\fid}[0]{\ensuremath{\mc{F}}\xspace}
\newcommand{\infid}[0]{\ensuremath{\varphi}\xspace}
\newcommand{\diam}[0]{\ensuremath{\eta}\xspace}
\newcommand{\leak}[0]{\ensuremath{\mc{L}}\xspace}
\newcommand{\vecsig}[0]{\ensuremath{\bm{\sigma}}\xspace}
\newcommand{\idop}[0]{\ensuremath{\op{\mathbb{1}}}\xspace}
\newcommand{\pid}[0]{\ensuremath{\op{P}_0}\xspace}
\newcommand{\px}[0]{\ensuremath{\op{P}_1}\xspace}
\newcommand{\py}[0]{\ensuremath{\op{P}_2}\xspace}
\newcommand{\pz}[0]{\ensuremath{\op{P}_3}\xspace}

\newcommand{\ud}[0]{\ensuremath{\uparrow \downarrow}\xspace}
\newcommand{\du}[0]{\ensuremath{\downarrow \uparrow}\xspace}

\newcommand{\T}[0]{\ensuremath{\intercal}\xspace}
\newcommand{\trace}[0]{\ensuremath{\mr{tr}}\xspace}
\newcommand{\cond}[0]{\ensuremath{\mr{cond}}\xspace}
\newcommand{\qp}{\ensuremath{\mc{E}}\xspace}
\newcommand{\db}[1]{\ensuremath{b_{#1}}\xspace}

\newcommand{\sts}[0]{$\mathrm{S\mbox{-}T_0}$\xspace}
\newcommand{\nseg}[0]{\ensuremath{N_{\mr{seg}}}\xspace}
\newcommand{\infidi}{\ensuremath{\infid_{\mr{initial}}}\xspace}
\newcommand{\infidf}{\ensuremath{\infid_{\mr{final}}}\xspace}

\newcommand{\vc}[1]{\ensuremath{\vec{#1}}\xspace}
\newcommand{\mat}[1]{\ensuremath{\mathbf{#1}}\xspace}
\newcommand{\op}[1]{\ensuremath{\hat{#1}}\xspace}

\newcommand{\refeq}[1]{Eq.\,\ref{#1}}
\newcommand{\reftab}[1]{Tab.\,\ref{#1}}
\newcommand{\reffig}[1]{Fig.\,\ref{#1}\,}
\newcommand{\refsec}[1]{Sec.\,\ref{#1}\,}

\newcommand{\g}[1]{%
    \IfEqCase{#1}{%
        {1}{\op{\mr{\textsc{cnot}}}}%
        {2}{\ensuremath{\op{\mr{\textsc{x}}}_{\pi/2}^{(1)}}}%
        {3}{\ensuremath{\op{\mr{\textsc{y}}}_{\pi/2}^{(1)}}}%
        {4}{\ensuremath{\op{\mr{\textsc{x}}}_{\pi/2}^{(2)}}}%
        {5}{\ensuremath{\op{\mr{\textsc{y}}}_{\pi/2}^{(2)}}}%
        {6}{\idop}
        {8}{}%
    }[\PackageError{g}{Undefined option to g: #1}{}]%
}%
\newcommand{\gshort}[1]{#1}%
\newcommand{\p}[3]{\ensuremath{p_{\gshort{#1},#2#3}}}
\newcommand{\m}[1]{%
    \IfEqCase{#1}{%
        {1}{\ensuremath{\op{\sigma}_{30}}}%
        {2}{\ensuremath{\op{\sigma}_{03}}}%
    }[\PackageError{m}{Undefined option to m: #1}{}]%
}%
\newcommand{\n}[1]{%
    \IfEqCase{#1}{%
        {qubits}{\ensuremath{n}\xspace}%
    }[\ensuremath{{n_{\mr{#1}}}}\xspace]%
}
\newcommand{\lasttablecaption}{}

\begin{document}

\title{Self-Consistent Calibration of Quantum Gate Sets}
\author{Pascal Cerfontaine\textsuperscript{1}}
\email[]{pascal.cerfontaine@rwth-aachen.de}
\author{Ren\'e Otten\textsuperscript{1}}
\author{Hendrik Bluhm\textsuperscript{1}}
\affiliation{\textsuperscript{1}JARA-FIT Institute for Quantum Information, Forschungszentrum J\"ulich GmbH and RWTH Aachen University, 52074 Aachen, Germany}

\date{April 28, 2020}

\begin{abstract}
The precise and automated calibration of quantum gates is a key requirement for building a reliable quantum computer. Unlike errors from decoherence, systematic errors can in principle be completely removed by tuning experimental parameters. Here, we present an iterative calibration routine which can remove systematic gate errors on several qubits. A central ingredient is the construction of pulse sequences that extract independent indicators for every linearly independent error generator. We show that decoherence errors only moderately degrade the achievable infidelity due to systematic errors. Furthermore, we investigate the convergence properties of our approach by performing simulations for a specific qubit encoded in a pair of spins. Our results indicate that a gate set with 230 gate parameters can be calibrated in about ten iterations, after which incoherent errors limit the gate fidelity.
\end{abstract}
\maketitle

\section{Introduction}
\label{sec:introduction}
The implementation of quantum gates with sufficiently low incoherent and coherent error rates is a key challenge for building reliable quantum computers. The elimination of coherent (systematic) errors is particularly pertinent as there is evidence that for the same average gate fidelity, coherent errors are more detrimental to error correction procedures than decoherence-induced errors \cite{Suzuki2017}.

A first step for the removal of such errors, which can often be achieved by appropriate changes to control pulse sequences, is to characterize them. In principle, process tomography extracts all relevant information about a single gate and can thus be used for this purpose. However, it is not very useful for initial gate tune-up as it requires fiducial states and measurements. These need to be either available with high accuracy or can be generated using gates -- which in turn need to be calibrated. The solution is to self-consistently characterize and calibrate a whole set of gates. Prominent self-consistent characterization protocols include Randomized Benchmarking (RB), Gate Set Tomography (GST) and Randomized Benchmarking Tomography \cite{Magesan2011, Blume-Kohout2017, Kimmel2014}. Even in the presence of decoherence as well as state-preparation and measurement (SPAM) errors, these methods can successfully extract information about gate errors. 

RB is based on repeatedly applying random Clifford gate sequences, which ideally retain the initial qubit state. For each gate sequence, the return probability to the initial state is measured. Since errors accumulate with increasing sequence length, an average error rate for the entire gate set can be extracted from an exponential fit of the return probability versus sequence length. In contrast, GST uses non-random gate sequences to obtain a full quantum process description of each gate. While RB lumps coherent and incoherent errors into a single figure, GST can thus distinguish between different error types. However, this comes at the cost of requiring more gate sequences and a much more involved self-consistent fit of all process parameters.
In addition to RB and GST, more specialized single-qubit approaches like Bootstrap Tomography \cite{Dobrovitski2010} or Robust Phase Estimation (RPE) \cite{Kimmel2015} have been proposed. Another single-qubit approach, AllXY \cite{Reed2013}, is specifically tailored to extract a subset of errors relevant to resonant control using a rotating wave approximation. All three methods extract more (but not necessarily complete) error information than RB but require fewer resources than GST.

Here, we also take a middle road between RB and GST. Our gate set characterization and calibration protocol (which we will from now on abbreviate as GSC) extends our previous work \cite{Cerfontaine2014, Cerfontaine2016, Cerfontaine2019ex} based on the bootstrap tomography approach from \citet{Dobrovitski2010} to larger gate sets and more than one qubit. It can be made self-consistent, or applied in a simplified version if single-qubit gates are already available. For \n{qubits} qubits, we focus on all $2^{2\n{qubits}}-1$ coherent errors generators of a gate (as opposed to $2^{4\n{qubits}} - 2^{2\n{qubits}}$ process parameters) since only these can in principle be completely removed by tuning control parameters. For a given set of available initial states, gates and measurements, GSC automatically constructs short gate sequences for the extraction of all coherent gate errors. Using these sequences in experiments is straightforward and requires much fewer measurements than GST and provides more detailed information than RB. Furthermore, almost no post-processing of the measurement results is required, decoherence errors only moderately degrade the achievable fidelity, and the sequences can also be made insensitive to SPAM errors.

Once gate errors have been characterized, they can be corrected by manual or automatic tuning of the available parameters. Automatic tuning is indispensable for scaling up to many qubits and to systematically deal with non-orthogonal gate parameters and cross-talk \cite{Cerfontaine2019ex, Cerfontaine2016, Frank2017, Cerfontaine2014, Egger2014, Kelly2014, Kelly2016}. The moderate resource requirements of GSC are especially advantageous for automated calibration, which feeds the extracted gate errors into an iterative optimization algorithm (from now on referred to as solver).

While GSC can be used in combination with different optimization algorithms, we focus on investigating the convergence properties of derivative-based solvers. As an example we investigate the calibration of a gate set for two encoded spin qubits \cite{Cerfontaine2019, Petta2005} since a less generic single-qubit version of our calibration routine has already successfully tuned single-qubit gates with 50 free parameters in this system \cite{Cerfontaine2019ex}. Our simulations demonstrate that a numerically optimized gate set with up to 230 free parameters can be calibrated. Based on a realistic noise model, this gate set is predicted to be decoherence-limited with a fidelity of around \SI{99.8}{\%}. Using GSC, fidelities beyond \SI{99.6}{\%} including decoherence and leakage can be reliably reached in about 10 iterations when tuning a \g1 gate with initial systematic infidelities as large as \SI{20}{\%}.

The paper is organized as follows: In \refsec{sec:gate_sequences}, we formally describe the general setting and algorithm which automatically constructs the gate sequences. Then we outline in \refsec{sec:tuning_algorithm} how these gate sequences can be used for tuning. We augment this by a discussion of possible error sources and the impact of decoherence on the calibration routine in \refsec{sec:error_sources}, and give explicit gate sequences for a two-qubit gate set in \refsec{sec:clifford_gate_set}. Last, we present our simulation of the calibration routine for two exchange-coupled spin qubits in \refsec{sec:application_to_st_qubits}. We mark vectors by arrows, use boldface for matrices, and set scalars in lightface.

\section{Construction of gate sequences}
\label{sec:gate_sequences}
The main idea of GSC is to initialize a well-defined state and apply a sequence of gates from the set of available gates, followed by a measurement. By repeating this process for different gate sequences and optionally initial states and measurements, all systematic gate errors can be extracted in a self-consistent manner. In this section we describe how GSC automatically generates a set of suitable gate sequences when the available initial states, gates and measurement operators are given.

We assume that the system consisting of \n{qubits} qubits with Hilbert space dimension $d = 2^{\n{qubits}}$ can be initialized in any of \n{i} initial states $\op{\rho}_i$ and read out using any of \n{m} measurements $\op{M}_m$. In order to calibrate a set of \n{g} quantum gates with unitary operators $\op{G}_g$, the algorithm first constructs all $\n{s} = \sum_{l=1}^{\n{l}} \n{g}^l$ gate sequences up to a certain length \n{l} with index $s = 1 ,\ldots, \n{s}$. Each sequence is a list of sequentially applied gates with length $l(s) \le \n{l}$. We define $\gamma_{s,t}$ as the index of the gate to be applied at time index $t = 1 ,\ldots, l(s)$ for sequence $s$. We assume that all gates can be applied independently and that cross-talk between simultaneously applied single qubit gates as well as concatenation errors are negligible. The unitary operator of sequence $s$ is then given by $\prod_{t=l(s)}^{1} \op{G}_{\gamma_{s,t}}$.

To assess which sequences are most suitable to extract gate errors, we first define the coherent error operator for gate $\op{G}_g$
\begin{align}
\op{E}_{g}(\vc{p}_g) = \prod_{k=1}^{d^2-1} \frac{ \idop - i p_{g,k} \op{\sigma}_k }{ \sqrt{1 + p_{g,k}^2} }
\label{eq:errop}
\end{align}
in terms of the $\n{p} = \n{g}(d^2-1)$ error parameters $p_{g,k}$, using the $d$-dimensional Pauli basis $\op{\sigma}_k \in (\pid, \px, \py, \pz)^{\otimes \n{qubits}}$ with $\pid = \idop$. The unitary operators describing each sequence perturbed by all possible unitary errors are then given by
\begin{align}
    \op{U}_{s}(\mat{p}) = \prod_{t=l(s)}^{1} \op{G}_{\gamma_{s,t}} \op{E}_{\gamma_{s,t}}(\vc{p}_{\gamma_{s,t}}),
    \label{eq:us}
\end{align}
where \mat{p} is the matrix representation of the error parameters $p_{g,k}$. We consider all $\n{r} = \n{i} \n{m} \n{s}$ measurement responses
\begin{align}
    R_{s,i,m}(\mat{p}) = \trace({\op{U}_{s}(\mat{p}) \op{\rho}_{i} \op{U}_{s}(\mat{p})^\dagger \op{M}_m}),
\end{align}
arising from applying \n{s} sequences to \n{i} different states followed by \n{m} different measurements.
In order to extract the measurement responses experimentally, each sequence needs to be applied to the respective initial state several times so that expectation values of the measurements can be computed. 

To simplify the notation, we now replace the double index $(g,k)$ with a single index $u = 1 ,\ldots, \n{p}$, and thus obtain the vector representation $\vc{p}$ of the matrix \mat{p}. Likewise, we write $R_{s,i,m}$ as a vector by replacing the triple index $(s,i,m)$ with a single index $r = 1 ,\ldots, \n{r}$. 

We assume small gate errors, and expand $\vc{R}(\vc{p})$ around $\vc{R}^{(0)} = \vc{R}(\vc{0})$,
\begin{align}
    R_{r}(\vc{p}) - R_r^{(0)}    & = \sum_{u=1}^{\n{p}} \frac{\partial  R_r}{\partial p_{u}} \Bigr|_{\vc{p}=0} \, p_{u} + \mathcal{O}(p_i p_j).
    \label{eq:m}
\end{align}
The matrix elements $S_{ru} = \partial R_r/\partial p_{u} |_{\vc{p}=0}$ can be calculated analytically or numerically using finite differences. If \mat{S} is square and invertible, and given the measurements $\vc{R}(\vc{p})$ and knowing the measurement value $\vc{R}^{(0)}$ theoretically expected for perfect gates, it is possible to solve for $\vc{p}$ and thus detect and calibrate all coherent gate errors. But as we will discuss in \refsec{sec:tuning_algorithm}, explicitly solving for $\vc{p}$ is not necessary for calibration purposes. Depending on the set of gates, initial states and measurements, it might not be possible to extract all coherent errors and \mat{S} will be singular. In this case, additional gates, states or measurements need to be considered.

If \mat{S} has more rows (sequences) than columns (coherent errors) and its rank is \n{p}, some of the sequences contain redundant information so that a subset of \n{p} rows of $\mat{S}$ can still have rank \n{p} and be invertible. As a criterion for invertibility we use that the condition number of $\mat{S}$ may not be pathologically large. A minimal subset of \n{p} linearly independent rows can be determined by using a rank-revealing QR decomposition \cite{Chan1987} on the transpose, $\mat{S}^\T \mat{P}_S = \mat{Q}_S \mat{R}_S$, with a unitary matrix $\mat{Q}_S$ and an upper triangular matrix $\mat{R}_S$. The permutation matrix $\mat{P}_S$ is chosen such that the absolute value of the diagonal elements of $\mat{R}_S$ is decreasing. If $\mat{S}$ has rank \n{p}, all diagonal elements of $\mat{R}_S$ will be nonzero. If this is the case, the algorithm uses the matrix $\mat{R}_{S,\min}$ containing only the first \n{p} columns (sequences) of $\mat{R}_S$. Choosing the columns with the largest diagonal elements in this manner ensures that errors can be extracted well. Next, the algorithm obtains $\mat{S}_{\min}^{\T} = \mat{Q}_{S} \mat{R}_{S,\min} \mat{P}_{S}^{\T}$. Using $\mat{S}_{\min}$, all coherent errors can be extracted from \vc{R} to first order if they are sufficiently small.

\section{Tuning algorithm}
\label{sec:tuning_algorithm}
In experiments, gates are controlled by some experimental parameters $q_j, j = 1 ,\ldots, \n{q}$, which affect the coherent error parameters \vc{p} via an unknown transformation. We do not make any assumption about the form of this transformation, we only assume that it is possible to find some set of experimental parameters $\vc{q}^{\,(0)}$ which remove all coherent gate errors.

Furthermore, realistic measurements can be affected by non-unitary dynamics and SPAM errors, so that we cannot directly extract $\vc{R}(\vc{q})$. Instead, we obtain the erroneous measurement results $\vc{\mc{R}}(\vc{q})$, which only reduce to $\vc{R}(\vc{q})$ for an ideal experiment. For now we assume $\vc{\mc{R}}^{(0)} = \vc{R}^{(0)}$ but we discuss the validity of this assumption under various realistic error sources in the next section. 

Under these assumptions, we can use $\Delta \vc{\mc{R}}(\vc{q}^{\,(0)}) = \vc{\mc{R}}(\vc{q}^{\,(0)}) - \vc{R}^{(0)} = 0$ to obtain the calibration procedure by solving the following optimization problem iteratively, using measurement results for $\Delta \vc{\mc{R}}(\vc{q})$:
\begin{align}
    \min_{\vc{q}}|\Delta \vc{\mc{R}}(\vc{q})|^2.
    \label{eq:min}
\end{align}
Since several solutions generally exist, we require that we start sufficiently close to the desired solution. This should be ensured by appropriate pre-characterization of the qubit system. 

Solving the optimization problem works well as long as a decrease in $|\Delta \vc{\mc{R}}(\vc{q})|$ leads to lower systematic errors, even if nonlinear terms in \refeq{eq:m} cannot be neglected. Various algorithms for calculating updates for $\vc{q}$ exist but since an optimal choice depends on the specific experimental implementation and number of free parameters, we will not go into detail here. For our simulations we choose the gradient-based Levenberg-Marquardt algorithm (LMA) \cite{Levenberg1944}, since it can take advantage of a vector-valued objective function. Since GSC is restricted to coherent errors, the number of required measurements just depends on the number of extracted parameters. Thus, it is experimentally feasible to obtain gradient information required by the LMA by measuring finite differences \cite{Cerfontaine2019ex}. While alternative approaches to estimate the Jacobian may turn out to be more efficient, this is beyond the scope of this work.

\section{Calibration errors}
\label{sec:error_sources}
So far, we have only considered coherent errors of the gates themselves, which GSC can calibrate in a self-consistent manner. However, decoherence, leakage out of the computational subspace and SPAM errors, or simply a faulty measurement calibration, could add offsets to each measurement.

Specifically, we assume that such errors can be modelled as offsets \vc{\delta} and visibility factors \vc{\gamma} in the measurement of sequence $r$, $\mc{R}_r(\vc{q}) = \gamma_r R_r(\vc{q}) - \delta_r$, where $\delta_r$ and $\gamma_r$ can depend on \vc{q}. Then, the calibration will null $\Delta \mc{R}_r(\vc{q}) = \gamma_r R_r(\vc{q}) - \delta_r - R_r^{(0)}$ so that $\gamma_r R_r(\vc{q}) -  R_r^{(0)} = \delta_r$ if the calibration has converged. In this case reaching $\vc{q} = \vc{q}^{\,(0)}$ is not guaranteed even if the calibration routine has converged. 

First, we discuss how coherent and incoherent gate and SPAM errors lead to offset and visibility errors, and how they affect different gate fidelity metrics. We also briefly comment on the effect of energy relaxation processes for single qubits. Throughout this section, we restrict ourselves to dynamics within the qubit subspace. This is justified for qubit systems for which the timescales of leakage processes are much larger than for decoherence processes.  We find that if decoherence is sufficiently low, the infidelity of gates after being tuned by GSC is not limited by remaining coherent errors. In addition, we give a recipe for the construction of gate sequences which are robust to offsets and coherent SPAM errors. Example gate sequences which implement the ideas presented in this section are presented in \refsec{sec:clifford_gate_set} for a two-qubit gate set.

\subsection{Stochastic errors}
\label{sec:unital}
We now discuss calibration errors caused by stochastic gate and SPAM errors in terms of two common fidelity metrics. For a completely positive trace-preserving quantum process \qp, which represents either an imperfect gate or a whole sequence, we consider the average gate infidelity from the identity $\infid(\qp) = 1-\fid(\qp, \idop)$ \cite{Nielsen2002}, and the diamond norm distance from the identity, $\diam(\qp)$ \cite{Magesan2012}. While the average infidelity is the most common metric, the diamond norm distance yields an error rate which can be directly compared to error correction thresholds.

For a superoperator \op{\qp} over the Hilbert space $\mathcal{H}$, the diamond norm is defined in terms of the trace norm as $||\op{\qp}||_{\lozenge} := \mr{sup}_{\mathcal{H}'} \mr{sup}_{\op{\rho}} ||\op{\qp} \otimes \idop(\op{\rho}) ||_\mr{Tr}$ where \op{\rho} is a density operator over the joint Hilbert space of the original system and an ancilla $\mathcal{H} \otimes \mathcal{H}'$ \cite{Wallman2016a}. Throughout this work, we use the diamond norm distance from the identity $\diam(\qp) = ||\op{\qp} - \idop ||_{\lozenge}$. 

We start by modelling coherent errors of a calibrated gate as $\qp_{\mr{c}}(\op{\rho}) = \op{E}_g(\vc{p}) \rho \op{E}_g^{\dagger}(\vc{p})$ using \refeq{eq:errop}. The contribution of the coherent error $p_k$ to the average gate infidelity can be calculated as $\infid_{\mr{c}} = \frac{d}{d+1} p_k^2$. Furthermore, the diamond distance scales linearly with $|p_k|$ and has an upper bound of $\diam_{\mr{c}} \le d |p_k|$ \cite{Wallman2016a}. If several small coherent errors $|\vc{p}|^2 \ll 1$ are present at the same time, $\infid_{\mr{c}} \approx \sum_k \frac{d}{d+1} p_k^2 = \frac{d}{d+1} |\vc{p}|^2$. In this case, the diamond distance has an upper bound of $\diam_{\mr{c}} \lessapprox d |\vc{p}|$.

As a model for decoherence, we consider a generalized Pauli channel \cite{Magesan2012} 
\begin{align}
    \qp_{\mr{i}}(\op{\rho}) = \sum_{k=0}^{d^2-1} a_k \op{\sigma}_k \op{\rho} \op{\sigma}_k^{\dagger}
\end{align}
describing stochastic (incoherent) errors \cite{Wallman2016}, where $\op{\sigma}_0 = \idop$. Since $\qp_{\mr{i}}$ is trace-preserving $\sum_k a_k \op{\sigma}_k \op{\sigma}_k^{\dagger} = \idop$, we know that $\sum_k a_k = 1$ with $a_k \ge 0$. Thus, we can interpret $a_0$ as the probability of retaining the original state \op{\rho}, and $a_k$ as the probability of an error of type $\op{\sigma}_k$ occurring. The infidelity of this channel is $\infid_{\mr{i}} = \frac{d}{d+1} (1 - a_0)$, while the diamond distance is given by $\diam_{\mr{i}} = 1 - a_0$ \cite{Magesan2012}. Thus, both measures are invariant if the operators $\op{\sigma}_k$ are replaced by $\op{U}^{\dagger} \op{\sigma}_k \op{U}$, resulting in the twirled channel $\qp_{\mr{i},U}$. This channel can be used to model several common noise sources such as dephasing and depolarization processes (but not relaxation and other non-unital processes). We limit the following discussion to $\qp_{\mr{i}}$ since fidelity and diamond distance are the same as for $\qp_{\mr{i},U}$, but will show that similar results can be obtained at least for special non-unital channels describing single-qubit relaxation. The decisive property of the error model is that it contains no unitary component, thus reflecting the situation once all systematic errors have been nulled by the calibration. 

We define $\qp_{\mr{i},r}$ in the form just described as the total decoherence error of each sequence $r$, arising from the corresponding errors of the gates in the sequence as well as possible SPAM errors. We can thus assume that the syndrome measurement at the end of the sequence is described by the ideal operator \op{M}. The worst-case syndrome measurement error is then given by $\max_{\op{\rho}} |\trace(\qp_{\mr{i},r}(\op{\rho}) \op{M}) - \trace(\op{\rho} \op{M})|$ and is upper bounded by $\lambda (1 - a_0) = \lambda \diam_{\mr{i},r} = \lambda \frac{d+1}{d} \infid_{\mr{i},r}$. Here, $\lambda = |\lambda_{\max} - \lambda_{\min}|$ is the difference between the maximal and minimal eigenvalues of \op{M}. For simplicity we restrict the following discussion to measurement operators \op{M} for which $\lambda = 1$. Then, the process $\qp_{\mr{i},r}$ with an infidelity $\infid_{\mr{i},r}$ or diamond distance $\diam_{\mr{i},r}$ leads to measurement errors $|\delta_r| \le \diam_{\mr{i},r} = \frac{d+1}{d} \infid_{\mr{i},r}$. 

Since the measurement errors \vc{\delta} will lead to residual coherent errors, even after the calibration has converged, we would now like to bound the effect on the infidelity $\infid_{\mr{c},g}$ and diamond distance $\diam_{\mr{c},g}$ of one gate due to coherent error parameters $\vc{p}_g$. For $|\vc{\delta}| \ll 1$ we can then use $\vc{R}(\vc{q}) -  \vc{R}^{(0)} = \vc{\delta}$ in \refeq{eq:m} to obtain the residual coherent errors of all gates $\vc{p} = \mat{S}^{-1} \vc{\delta} + \mathcal{O}(\delta_i \delta_j)$. With $b = \frac{d}{d+1}$ we obtain
\begin{align}
    \infid_{\mr{c},g} & = b \left|\vc{p}_g\right|^2 \le b \left|\vc{p}\right|^2 = b \left|\mat{S}^{-1} \vc{\delta}\right|^2 \label{eq:tighter_error_bound} \\
    & \le b \left|\mat{S}^{-1}\right|^2 \left|\vc{\delta}\right|^2 \le \left|\mat{S}^{-1}\right|^2  \left|\vc{\infid}_{\mr{i}}\right|^2,
    \label{eq:error_bound}
\end{align}
where $|\mat{S}^{-1}|$ is the Frobenius norm of $\mat{S}^{-1}$. Therefore the infidelity $\infid_{\mr{c},g}$ of gate $g$ from residual coherent errors is bounded quadratically by the norm of the sequence infidelities $\vc{\infid}_{\mr{i}}$ from stochastic errors. This is very favourable as long as $\left|\vc{\infid}_{\mr{i}}\right| < 1$. If we repeat the same analysis  for the diamond distance, we get
\begin{align}
    \diam_{\mr{c},g} & \lessapprox d \left|\vc{p}_g\right| \le d \left|\vc{p}\right| = d \left|\mat{S}^{-1} \vc{\delta}\right| \label{eq:tighter_error_bound_diam} \\
    & \le d \left|\mat{S}^{-1}\right| \left|\vc{\delta}\right| \le d \left|\mat{S}^{-1}\right|  \left|\vc{\diam}_{\mr{i}}\right|,
    \label{eq:error_bound_diam}
\end{align}
Thus, the diamond distance $\diam_{\mr{c},g}$ of gate $g$ due to residual coherent errors is bounded linearly by the norm of the sequence diamond distances $\vc{\diam}_{\mr{i}}$ from stochastic errors. For both fidelity metrics the error contribution from residual coherent errors is bounded by the total error contribution from stochastic errors. While the scaling is more favourable for the infidelity, the error is still well-controlled when using the diamond distance.

Note that $\infid_{\mr{i},r}$ and $\diam_{\mr{i},r}$ reflect the total decoherence error of a sequence. If the underlying noise is Markovian on the time scale of the duration of each gate, these metrics are the sum of those of $l(r)$ individual gates to lowest order. Thus, they are roughly a factor $l(r)$ larger than the typical gate error metrics. If correlations leading to constructive interference of errors exist, a less favorable scaling with $l(r)^2$ may be encountered.

Naturally, the bounds need not be saturated. A numerical prefactor that is smaller than $|\mat{S}^{-1}|$ can be obtained by simply considering $\vc{p}_g$ rather than all of $\vc{p}$ in the above inequalities. Moreover, not all decoherence processes affect the measurement outcomes in the same way. As an extreme example, let us assume that in the absence of coherent errors all sequences prepare a single-qubit state on the equator of the Bloch sphere. If the decoherence of each sequence $\qp_{\mr{i},r}$ can be modelled as a dephasing channel around the z-axis, a measurement of \pz will be unaffected by $\qp_{\mr{i},r}$ and thus result in $\vc{\delta} = \vc{0}$. This example shows that it may be useful to search for sets of state and measurements for which the decoherence process does not change the measurement outcomes to design sequences with tighter bounds. 

In addition, the bounds can be improved by reducing $\left|\mat{S}^{-1}\right|$, i.e. by constructing longer gate sequences which contain the same gate $N$ times to accumulate coherent errors. Thus, some elements of $\mat{S}$ will grow but by no more than a factor $N$. However, additional gates will generally also increase $\vc{\infid}_{\mr{i}}$ so that longer sequences are only advantageous if some part of $\vc{\infid}_{\mr{i}}$ grows slower than $N$. Again, this might be due to a decoherence process which does not change the measurement outcomes. Even if both the relevant matrix elements of \mat{S} and $\vc{\infid}_{\mr{i}}$ grow linearly with $N$, longer gate sequences still have the benefit of reducing the sensitivity to SPAM errors and increasing the readout signal. Hence, fewer measurements are needed to calculate expectation values without increasing the upper fidelity bounds of \refeq{eq:error_bound} and \refeq{eq:error_bound_diam}.

We would like to point out that the above arguments for stochastic errors can also be applied to single-qubit energy relaxation processes. Energy relaxation is a common non-unital error encountered in all qubit systems, typically associated with an exponential decay with the characteristic timescale $T_1$. The associated infidelity is $\infid_{T_1} = \frac{d}{d+1} \frac{t}{2T_1} + \mathcal{O}(\sfrac{t^2}{T_1^2})$, where the qubit is subjected to energy relaxation for a time $t$. We numerically determine the diamond distance for single-qubit $T_1$ processes \cite{Watrous2009,daSilva2014}, and obtain a linear scaling with $\frac{t}{T_1}$. Thus both \infid and \diam scale linearly with $\frac{t}{T_1}$, the same scaling we used for the discussion of stochastic errors.

\subsection{Faulty readout calibration and coherent SPAM errors}
\label{sec:calib_spam}
In this section we examine the effects of coherent SPAM errors and a faulty readout calibration, and outline possible approaches to increase the robustness of GSC with respect to these errors.

Coherent SPAM errors can be considered by introducing an additional gate and error parameters for each measurement and readout. We define $\op{G}_g = \idop$ and thus introduce additional $p_{g,k}$, where $g = \n{g}+1 ,\ldots, \n{g}+\n{i}$ for initialization errors and $g = \n{g}+\n{i}+1 ,\ldots, \n{g}+\n{i}+\n{m}$ for measurement errors. These additional gates are treated exactly like the regular gates except that they are only applied once in each sequences before (after) the corresponding measurement (initialization). This approach overparametrizes the initialization errors by considering their effect on all possible initial states, but treating them on the same footing as gate errors is very convenient computationally. Using this method we define the modified unitary sequence operators $\op{V}_{s,i,m}$ and modified measurement results $R_{s,i,m}$:
\begin{align}    
    \op{V}_{s,i,m}(\vc{p}) & = \op{G}_{\n{g}+\n{i}+m} \op{U}_{s}(\vc{p}) \op{G}_{\n{g}+i} \\
    R_{s,i,m}(\vc{p}) & = \trace(\op{V}_{s,i,m}(\vc{p}) \op{\rho}_{i} \op{V}_{s,i,m}(\vc{p})^\dagger \op{M}_m).
\end{align}
Given specific gate sequences, we can then calculate which coherent SPAM errors enter to first or only to second order in \vc{\delta}. There will always be some SPAM errors which enter to first order, otherwise GSC would not be first-order sensitive to gate errors. 

Remember that a key requirement of our protocol is to require full rank with respect to the first $\n{g}(d^2-1)$ gate error parameters when selecting a subset of rows (sequences) of $\mat{S}$. In addition, we can enforce first-order insensitivity to some of the SPAM error parameters represented by gates with index $g > \n{g}$. It is possible to construct gate sequence with no first-order sensitivity to almost all single-qubit SPAM errors as demonstrated for a two-qubit gateset in \refsec{sec:clifford_gate_set}. To what extent this reduction of the number of relevant error generators leads to a higher accuracy depends on the specific errors present in the gate set to be calibrated.

Next, we consider the effect of a faulty readout calibration, including both offsets \vc{\delta} and visibility factors \vc{\gamma}. For known \vc{\delta}, the bounds are directly given by \refeq{eq:tighter_error_bound} and \refeq{eq:tighter_error_bound_diam}. For the measurement result $\mc{R}_r(\vc{q}) = \gamma_r R_r(\vc{q}) - \delta_r$ the symmetric visibility loss $\gamma_r$ does not lead to a biased calibration if $R_r^{(0)} = 0$. This condition can be easily enforced in the sequence construction. All sequences presented in this work obey this condition, except the last two rows in \reftab{tab:reg_seq_cnot_sim} as discussed later in \refsec{sec:clifford_gate_set}. 

In order to make the calibration insensitive to offset errors \vc{\delta}, we can construct a second complementary sequence for each of the original sequences. The resulting complementary measurement results $\Delta \vc{R}^{(2)}$ are then subtracted from the original measurement results $\Delta \vc{R}^{(1)}$ yielding $\Delta \vc{R} = \Delta \vc{R}^{(1)} - \Delta \vc{R}^{(2)} $. This will cancel out any additive errors that are the same for each pair, such as readout calibration errors. The constructions works as follows: In order to be still able to detect and calibrate all coherent gate errors we first require that the difference of the sensitivity matrices $\mat{S}^{(1)} - \mat{S}^{(2)}$ is invertible with good condition number. Each sequence pair should use the same measurement, so that the minimal number of additional sequences needed is at least \n{m}, and possibly more to ensure invertibility of $\mat{S}^{(1)} - \mat{S}^{(2)}$. Note that the number of complementary sequences can be lower than the number of primary sequences by paring the same complementary sequence with multiple primary sequences. 

\subsection{Cross-talk}
In this section, we briefly comment on the effect of classical and non-classical cross-talk on the calibration.

A straightforward approach within the GSC framework would be to treat each combination of single-qubit gates executed in parallel as a two-qubit gate, thus introducing additional coherent error parameters. To reduce the number of additional parameters, we can limit the number of single-qubit gate pairs (e.g. only allow operations on one qubit while the other performs an identity). Since each pair introduces new systematic errors, more sequences will generally be required.

If the additional parameters are not explicitly considered by adding more sequences or if there are no gate parameters which can be adjusted to address them, this will generally lead to inconsistencies of the measurement results for the various sequences. The magnitude of the resulting measurement errors \vc{\delta} will be given by the cross-talk strength. For measurement errors \vc{\delta}, the arguments from \refsec{sec:unital} can be applied to bound the effect on the achievable fidelity and diamond distance.
  
\section{Two-qubit gate set}
\label{sec:clifford_gate_set}
In this section, we give explicit gate sequences for calibrating a two-qubit gate set, implementing the ideas from \refsec{sec:calib_spam}. Specifically, we consider the gate set $\op{G} \in \left\{ \g1, \g2, \g3, \g4, \g5 \right\}$, where superscripts indicate on which qubit the gate is executed. To represent coherent gate errors, we use the basis elements $\op{P}_i \otimes \op{P}_j$ and corresponding error strengths $p_{g,ij}$ in \refeq{eq:errop}. The initial state is $\ket{00}$ and we consider the measurements $\m1 = \op{P}_3 \otimes \op{P}_0$ and $\m2 = \op{P}_0 \otimes \op{P}_3$. We denote coherent initialization and measurement errors by $p_{\mr{i},ij}$ and $p_{\mr{m},ij}$, where $p_{\mr{m},ij}$ is understood to refer \m1 or \m2, whichever is used in the sequence. 

For all of the gate sequence presented in this section, $|\mat{S}^{-1}|$ takes on values between $5.1$ and $8.1$. The achievable condition number depends on the number of additional constraints and varies between 8.4 and 83.5 for the examples discussed in this section. If the gate sequences are not chosen carefully, the condition number can deviate from the optimum by a few percent up to infinity if the system of equations is not invertible.

If single-qubit gates are already calibrated, i.e. by running a single-qubit version of GSC on each qubit, the \g1 gate can be tuned by using the sequences given in \reftab{tab:reg_seq_cnot_sim}. Each row contains the gate sequence, measurement operator and expected measurement result with the linear dependence on the coherent gate error parameters $p_{g,ij}$. Higher order terms are omitted.

Notice that all sequences are designed to produce the same measurement statistics as a completely mixed state. While our intention is to start the optimization close to a solution, this might not be the case. In order to make sure that the initial iterations do not run towards gates that produce the completely mixed state, we can introduce additional sequences. In this case, we give sequences 16 and 17 as an example, which ideally reproduce the initial state. Once the calibration is close to the desired target, they can be omitted from the calibration. Note that a sensible choice of such additional sequences requires some insight into the decoherence mechanisms, since they need to cover all types of excess decoherence that can occur in the gate set at hand.

By using the approach outlined in \refsec{sec:calib_spam}, we also generate sequences for tuning the \g1 gate that are first-order insensitive to all coherent single-qubit initialization errors, and first-order insensitive to all coherent single-qubit measurement errors except $\propto \idop \otimes \op{\sigma}_x$, shown in \reftab{tab:insens_seq_cnot}. This error corresponds to a single-qubit rotation around the $x$-axis directly before the perfect readout. This result implies that only a single type of coherent error of the readout operation needs to be minimized to reduce SPAM errors.

If single-qubit gates are to be tuned simultaneously, the sequences in \reftab{tab:diff_seq_all} can be used. Adding 10 sequences for 12 new single-qubit error parameters is sufficient since \pz measurements only determine the x- and y-axis of each qubit up to z-rotations. Thus, we obtain the condition number of $83.5$ by adding two equations setting the y-component of the rotation axis of \g2 and \g4 to zero. Four additional sequences to control for decoherence of each single-qubit gate are not listed explicitly, but can be generated in the same manner as before by repeating each gate twice and measuring the appropriate qubit. Sequences 26 and 27 are designed for dealing with energy relaxation, leakage and offsets in the measurement calibration (in the appendix we also provide more sequences for dealing with offsets).

\begin{table*}

\centering
\renewcommand\arraystretch{1.3}
\begin{tabular}{r *{8}{l}}
\hline\hline
s & $\gamma_{s,1}$ & $\gamma_{s,2}$ & $\gamma_{s,3}$ & $\gamma_{s,4}$ & $\op{M}_s$ & $R_r(\frac{1}{2} \vc{p})$ for Gates & Initialization & Measurement \\
\hline
1 & \g1~ & \g2~ & \g8~ & \g8~ & \m1~ & $0-\p{1}{1}{1} +\p{1}{2}{2} $ & $-\p{\mathrm{i}}{1}{1} +\p{\mathrm{i}}{2}{2}  $ & $-\p{\mathrm{m}}{1}{0} -\p{\mathrm{m}}{1}{3}  $ \\
2 & \g2~ & \g1~ & \g8~ & \g8~ & \m1~ & $0-\p{1}{1}{0} -\p{1}{1}{3} $ & $-\p{\mathrm{i}}{1}{0} -\p{\mathrm{i}}{1}{3}  $ & $-\p{\mathrm{m}}{1}{1} +\p{\mathrm{m}}{2}{2}  $ \\
3 & \g1~ & \g3~ & \g8~ & \g8~ & \m1~ & $0-\p{1}{2}{1} -\p{1}{1}{2} $ & $-\p{\mathrm{i}}{2}{1} -\p{\mathrm{i}}{1}{2}  $ & $-\p{\mathrm{m}}{2}{0} -\p{\mathrm{m}}{2}{3}  $ \\
4 & \g3~ & \g1~ & \g8~ & \g8~ & \m1~ & $0-\p{1}{2}{0} -\p{1}{2}{3} $ & $-\p{\mathrm{i}}{2}{0} -\p{\mathrm{i}}{2}{3}  $ & $-\p{\mathrm{m}}{2}{1} -\p{\mathrm{m}}{1}{2}  $ \\
5 & \g1~ & \g4~ & \g8~ & \g8~ & \m2~ & $0-\p{1}{0}{1} -\p{1}{3}{1} $ & $-\p{\mathrm{i}}{0}{1} -\p{\mathrm{i}}{3}{1}  $ & $-\p{\mathrm{m}}{0}{1} -\p{\mathrm{m}}{3}{1}  $ \\
6 & \g1~ & \g5~ & \g8~ & \g8~ & \m2~ & $0-\p{1}{0}{2} -\p{1}{3}{2} $ & $-\p{\mathrm{i}}{0}{2} -\p{\mathrm{i}}{3}{2}  $ & $-\p{\mathrm{m}}{0}{2} -\p{\mathrm{m}}{3}{2}  $ \\
7 & \g2~ & \g1~ & \g2~ & \g8~ & \m1~ & $0-\p{1}{0}{2} +\p{1}{2}{2} $ & $-\p{\mathrm{i}}{0}{2} -\p{\mathrm{i}}{3}{2}  $ & $+\p{\mathrm{m}}{2}{2} -\p{\mathrm{m}}{1}{3}  $ \\
8 & \g2~ & \g1~ & \g3~ & \g8~ & \m1~ & $0-\p{1}{3}{1} -\p{1}{1}{2} $ & $-\p{\mathrm{i}}{2}{1} -\p{\mathrm{i}}{1}{2}  $ & $-\p{\mathrm{m}}{1}{1} -\p{\mathrm{m}}{2}{3}  $ \\
9 & \g3~ & \g1~ & \g2~ & \g8~ & \m1~ & $0+\p{1}{3}{1} +\p{1}{2}{2} $ & $-\p{\mathrm{i}}{1}{1} +\p{\mathrm{i}}{2}{2}  $ & $-\p{\mathrm{m}}{2}{1} -\p{\mathrm{m}}{1}{3}  $ \\
10 & \g2~ & \g4~ & \g1~ & \g8~ & \m2~ & $0-\p{1}{3}{1} -\p{1}{1}{3} $ & $-\p{\mathrm{i}}{2}{1} -\p{\mathrm{i}}{1}{2}  $ & $-\p{\mathrm{m}}{3}{1} +\p{\mathrm{m}}{2}{2}  $ \\
11 & \g3~ & \g1~ & \g3~ & \g8~ & \m1~ & $0-\p{1}{0}{2} -\p{1}{1}{2} $ & $-\p{\mathrm{i}}{0}{2} -\p{\mathrm{i}}{3}{2}  $ & $-\p{\mathrm{m}}{1}{2} -\p{\mathrm{m}}{2}{3}  $ \\
12 & \g3~ & \g5~ & \g1~ & \g8~ & \m2~ & $0-\p{1}{3}{2} -\p{1}{2}{3} $ & $+\p{\mathrm{i}}{2}{1} +\p{\mathrm{i}}{1}{2}  $ & $-\p{\mathrm{m}}{0}{2} -\p{\mathrm{m}}{1}{2}  $ \\
13 & \g5~ & \g1~ & \g4~ & \g8~ & \m2~ & $0+\p{1}{0}{3} +\p{1}{3}{3} $ & $-\p{\mathrm{i}}{0}{1} -\p{\mathrm{i}}{3}{1}  $ & $-\p{\mathrm{m}}{0}{2} -\p{\mathrm{m}}{3}{2}  $ \\
14 & \g2~ & \g1~ & \g1~ & \g3~ & \m1~ & $0-2\p{1}{3}{0} -\p{1}{0}{3} -\p{1}{3}{3} $ & $-\p{\mathrm{i}}{2}{0} -\p{\mathrm{i}}{2}{3}  $ & $-\p{\mathrm{m}}{1}{0} -\p{\mathrm{m}}{1}{3}  $ \\
15 & \g2~ & \g5~ & \g1~ & \g4~ & \m2~ & $0-\p{1}{1}{2} +\p{1}{3}{3} $ & $-\p{\mathrm{i}}{2}{1} -\p{\mathrm{i}}{1}{2}  $ & $-\p{\mathrm{m}}{0}{2} +\p{\mathrm{m}}{2}{2}  $ \\
16 & \g1~ & \g1~ & \g8~ & \g8~ & \m1~ & $1$ & $ $ & $ $ \\
17 & \g1~ & \g1~ & \g8~ & \g8~ & \m2~ & $1$ & $ $ & $ $ \\
\hline\hline
\end{tabular}

\renewcommand{\lasttablecaption}{8.4}

\caption{Full list of sequences for calibration of a \g1 gate using the basis elements $\op{P}_i \otimes \op{P}_j$ and corresponding error strengths \p{1}{i}{j}. Each row contains a gate sequence, the measurement operator ($\m1 = \op{P}_3 \otimes \op{P}_0$ or $\m2 = \op{P}_0 \otimes \op{P}_3$) and the expected linear dependence of the measurement outcome on the coherent error parameters. For clarity, the prefactor of each coefficient \p{1}{i}{j} was divided by 2. Sequences 1-15 are used for eliminating systematic errors while sequences 16 and 17 can be used to extract information on decoherence (and leakage). Sequences are applied from left to right. All sequences assume initialization in the $\ket{00}$ state. The condition number of the sensitivity matrix \mat{S} is $\lasttablecaption$.}
\label{tab:reg_seq_cnot_sim}
\end{table*}

\begin{turnpage}
\begin{table*}

\centering
\renewcommand\arraystretch{1.3}
\begin{tabular}{r *{12}{l}}
\hline\hline
s & $\gamma_{s,1}$ & $\gamma_{s,2}$ & $\gamma_{s,3}$ & $\gamma_{s,4}$ & $\gamma_{s,5}$ & $\gamma_{s,6}$ & $\gamma_{s,7}$ & $\gamma_{s,8}$ & $\op{M}_s$ & $R_r(\frac{1}{2} \vc{p})$ for Gates & Initialization & Measurement \\
\hline
1 & \g2~ & \g1~ & \g4~ & \g5~ & \g1~ & \g3~ & \g8~ & \g8~ & \m2~ & $0+\p{1}{1}{0} -\p{1}{0}{1} +\p{1}{1}{1} -\p{1}{3}{2} $ & $+\p{\mathrm{i}}{1}{1} -\p{\mathrm{i}}{2}{2}  $ & $-\p{\mathrm{m}}{0}{1} -\p{\mathrm{m}}{3}{1}  $ \\
2 & \g2~ & \g4~ & \g1~ & \g5~ & \g1~ & \g3~ & \g8~ & \g8~ & \m2~ & $0+\p{1}{1}{0} -\p{1}{0}{1} +\p{1}{1}{1} -\p{1}{3}{3} $ & $+\p{\mathrm{i}}{1}{1} -\p{\mathrm{i}}{2}{2}  $ & $-\p{\mathrm{m}}{0}{1} -\p{\mathrm{m}}{3}{1}  $ \\
3 & \g2~ & \g5~ & \g1~ & \g2~ & \g5~ & \g4~ & \g8~ & \g8~ & \m2~ & $0-\p{1}{1}{2} +\p{1}{3}{3} $ & $-\p{\mathrm{i}}{2}{1} -\p{\mathrm{i}}{1}{2}  $ & $+\p{\mathrm{m}}{0}{1} -\p{\mathrm{m}}{3}{1}  $ \\
4 & \g3~ & \g4~ & \g1~ & \g5~ & \g1~ & \g2~ & \g8~ & \g8~ & \m2~ & $0+\p{1}{2}{0} -\p{1}{0}{1} +\p{1}{2}{1} -\p{1}{3}{3} $ & $+\p{\mathrm{i}}{2}{1} +\p{\mathrm{i}}{1}{2}  $ & $-\p{\mathrm{m}}{0}{1} +\p{\mathrm{m}}{3}{1}  $ \\
5 & \g3~ & \g5~ & \g1~ & \g3~ & \g5~ & \g4~ & \g8~ & \g8~ & \m2~ & $0-\p{1}{2}{2} +\p{1}{3}{3} $ & $+\p{\mathrm{i}}{1}{1} -\p{\mathrm{i}}{2}{2}  $ & $+\p{\mathrm{m}}{0}{1} -\p{\mathrm{m}}{3}{1}  $ \\
6 & \g1~ & \g2~ & \g1~ & \g3~ & \g5~ & \g1~ & \g3~ & \g8~ & \m2~ & $0+\p{1}{1}{0} +\p{1}{3}{0} +\p{1}{0}{1} +\p{1}{2}{1} +\p{1}{1}{2} +\p{1}{3}{3} $ & $+\p{\mathrm{i}}{2}{1} +\p{\mathrm{i}}{1}{2}  $ & $+\p{\mathrm{m}}{0}{1} -\p{\mathrm{m}}{3}{1}  $ \\
7 & \g1~ & \g3~ & \g1~ & \g2~ & \g5~ & \g1~ & \g2~ & \g8~ & \m2~ & $0+\p{1}{2}{0} -\p{1}{3}{0} -\p{1}{0}{1} +\p{1}{1}{1} -\p{1}{2}{2} -\p{1}{3}{3} $ & $+\p{\mathrm{i}}{1}{1} -\p{\mathrm{i}}{2}{2}  $ & $-\p{\mathrm{m}}{0}{1} +\p{\mathrm{m}}{3}{1}  $ \\
8 & \g2~ & \g4~ & \g1~ & \g5~ & \g1~ & \g3~ & \g5~ & \g8~ & \m2~ & $0+\p{1}{3}{1} -\p{1}{2}{2} +\p{1}{1}{3} -\p{1}{3}{3} $ & $+\p{\mathrm{i}}{2}{1} +\p{\mathrm{i}}{1}{2}  $ & $-\p{\mathrm{m}}{0}{1} -\p{\mathrm{m}}{3}{1}  $ \\
9 & \g3~ & \g1~ & \g4~ & \g5~ & \g1~ & \g2~ & \g5~ & \g8~ & \m2~ & $0+\p{1}{3}{1} +\p{1}{1}{2} +\p{1}{2}{2} -\p{1}{3}{3} $ & $-\p{\mathrm{i}}{1}{1} +\p{\mathrm{i}}{2}{2}  $ & $-\p{\mathrm{m}}{0}{1} +\p{\mathrm{m}}{3}{1}  $ \\
10 & \g3~ & \g4~ & \g1~ & \g5~ & \g1~ & \g2~ & \g5~ & \g8~ & \m2~ & $0+\p{1}{3}{1} +\p{1}{1}{2} +\p{1}{2}{3} -\p{1}{3}{3} $ & $-\p{\mathrm{i}}{1}{1} +\p{\mathrm{i}}{2}{2}  $ & $-\p{\mathrm{m}}{0}{1} +\p{\mathrm{m}}{3}{1}  $ \\
11 & \g3~ & \g5~ & \g1~ & \g3~ & \g1~ & \g5~ & \g4~ & \g8~ & \m2~ & $0+\p{1}{2}{2} -\p{1}{0}{3} $ & $-\p{\mathrm{i}}{1}{1} +\p{\mathrm{i}}{2}{2}  $ & $+\p{\mathrm{m}}{0}{1} -\p{\mathrm{m}}{3}{1}  $ \\
12 & \g5~ & \g1~ & \g3~ & \g1~ & \g3~ & \g5~ & \g4~ & \g8~ & \m2~ & $0+\p{1}{1}{3} +\p{1}{3}{3} $ & $-\p{\mathrm{i}}{1}{1} +\p{\mathrm{i}}{2}{2}  $ & $+\p{\mathrm{m}}{0}{1} -\p{\mathrm{m}}{3}{1}  $ \\
13 & \g2~ & \g1~ & \g4~ & \g3~ & \g5~ & \g3~ & \g1~ & \g3~ & \m2~ & $0+\p{1}{1}{0} +\p{1}{0}{1} -\p{1}{1}{1} +\p{1}{3}{2} $ & $-\p{\mathrm{i}}{1}{1} +\p{\mathrm{i}}{2}{2}  $ & $+\p{\mathrm{m}}{0}{1} +\p{\mathrm{m}}{1}{1}  $ \\
14 & \g2~ & \g5~ & \g1~ & \g2~ & \g5~ & \g1~ & \g4~ & \g5~ & \m2~ & $0+\p{1}{0}{2} +\p{1}{1}{3} $ & $+\p{\mathrm{i}}{2}{1} +\p{\mathrm{i}}{1}{2}  $ & $-\p{\mathrm{m}}{0}{1} +\p{\mathrm{m}}{3}{1}  $ \\
15 & \g3~ & \g4~ & \g1~ & \g5~ & \g2~ & \g2~ & \g1~ & \g2~ & \m2~ & $0+\p{1}{2}{0} +\p{1}{0}{1} -\p{1}{2}{1} +\p{1}{3}{3} $ & $-\p{\mathrm{i}}{2}{1} -\p{\mathrm{i}}{1}{2}  $ & $+\p{\mathrm{m}}{0}{1} +\p{\mathrm{m}}{2}{1}  $ \\
\hline\hline
\end{tabular}

\renewcommand{\lasttablecaption}{11.5}

\caption{Gate sequences for calibration of a \g1 gate with $\cond(\mat{S}) = \lasttablecaption$. These sequences are insensitive to all coherent single-qubit errors during initialization, and all coherent single-qubit measurement errors except $\idop \otimes \op{\sigma}_x$. For example, sequence 3 works by preparing both qubits in superposition states before applying the \g1 gate. Even without the \g1 gate, state preparation errors affecting only the first qubit would not enter to first order. However, the sequence would be sensitive to state preparation errors affecting the second qubit. The \g1 gate flips the second qubit if the first qubit is in state $\ket{1}$, i.e. only for half of the terms of the superposition state. Thus, for the measurement \m2 the sign on these state preparation errors is flipped for half of the state, leading to the observed linear insensitivity.}
\label{tab:insens_seq_cnot}
\end{table*}
\end{turnpage}

\begin{table*}

\centering
\renewcommand\arraystretch{1.3}
\begin{tabular}{r *{8}{l}}
\hline\hline
s & $\gamma_{s,1}$ & $\gamma_{s,2}$ & $\gamma_{s,3}$ & $\gamma_{s,4}$ & $\op{M}_s$ & $R_r(\frac{1}{2} \vc{p})$ for Gates & Initialization & Measurement \\
\hline
1 & \g2~ & \g8~ & \g8~ & \g8~ & \m1~ & $0-\p{2}{1}{0} $ & $-\p{\mathrm{i}}{1}{0} -\p{\mathrm{i}}{1}{3}  $ & $-\p{\mathrm{m}}{1}{0} -\p{\mathrm{m}}{1}{3}  $ \\
2 & \g3~ & \g8~ & \g8~ & \g8~ & \m1~ & $0-\p{3}{2}{0} $ & $-\p{\mathrm{i}}{2}{0} -\p{\mathrm{i}}{2}{3}  $ & $-\p{\mathrm{m}}{2}{0} -\p{\mathrm{m}}{2}{3}  $ \\
3 & \g4~ & \g8~ & \g8~ & \g8~ & \m2~ & $0-\p{4}{0}{1} $ & $-\p{\mathrm{i}}{0}{1} -\p{\mathrm{i}}{3}{1}  $ & $-\p{\mathrm{m}}{0}{1} -\p{\mathrm{m}}{3}{1}  $ \\
4 & \g5~ & \g8~ & \g8~ & \g8~ & \m2~ & $0-\p{5}{0}{2} $ & $-\p{\mathrm{i}}{0}{2} -\p{\mathrm{i}}{3}{2}  $ & $-\p{\mathrm{m}}{0}{2} -\p{\mathrm{m}}{3}{2}  $ \\
5 & \g1~ & \g2~ & \g8~ & \g8~ & \m1~ & $0-\p{1}{1}{1} +\p{1}{2}{2} -\p{2}{1}{0} $ & $-\p{\mathrm{i}}{1}{1} +\p{\mathrm{i}}{2}{2}  $ & $-\p{\mathrm{m}}{1}{0} -\p{\mathrm{m}}{1}{3}  $ \\
6 & \g1~ & \g3~ & \g8~ & \g8~ & \m1~ & $0-\p{1}{2}{1} -\p{1}{1}{2} -\p{3}{2}{0} $ & $-\p{\mathrm{i}}{2}{1} -\p{\mathrm{i}}{1}{2}  $ & $-\p{\mathrm{m}}{2}{0} -\p{\mathrm{m}}{2}{3}  $ \\
7 & \g1~ & \g4~ & \g8~ & \g8~ & \m2~ & $0-\p{1}{0}{1} -\p{1}{3}{1} -\p{4}{0}{1} $ & $-\p{\mathrm{i}}{0}{1} -\p{\mathrm{i}}{3}{1}  $ & $-\p{\mathrm{m}}{0}{1} -\p{\mathrm{m}}{3}{1}  $ \\
8 & \g1~ & \g5~ & \g8~ & \g8~ & \m2~ & $0-\p{1}{0}{2} -\p{1}{3}{2} -\p{5}{0}{2} $ & $-\p{\mathrm{i}}{0}{2} -\p{\mathrm{i}}{3}{2}  $ & $-\p{\mathrm{m}}{0}{2} -\p{\mathrm{m}}{3}{2}  $ \\
9 & \g2~ & \g1~ & \g8~ & \g8~ & \m1~ & $0-\p{1}{1}{0} -\p{1}{1}{3} -\p{2}{1}{0} $ & $-\p{\mathrm{i}}{1}{0} -\p{\mathrm{i}}{1}{3}  $ & $-\p{\mathrm{m}}{1}{1} +\p{\mathrm{m}}{2}{2}  $ \\
10 & \g2~ & \g3~ & \g8~ & \g8~ & \m1~ & $0-\p{2}{2}{0} -\p{3}{3}{0} $ & $-\p{\mathrm{i}}{2}{0} -\p{\mathrm{i}}{2}{3}  $ & $-\p{\mathrm{m}}{1}{0} -\p{\mathrm{m}}{1}{3}  $ \\
11 & \g3~ & \g1~ & \g8~ & \g8~ & \m1~ & $0-\p{1}{2}{0} -\p{1}{2}{3} -\p{3}{2}{0} $ & $-\p{\mathrm{i}}{2}{0} -\p{\mathrm{i}}{2}{3}  $ & $-\p{\mathrm{m}}{2}{1} -\p{\mathrm{m}}{1}{2}  $ \\
12 & \g3~ & \g2~ & \g8~ & \g8~ & \m1~ & $0+\p{2}{3}{0} -\p{3}{1}{0} $ & $-\p{\mathrm{i}}{1}{0} -\p{\mathrm{i}}{1}{3}  $ & $-\p{\mathrm{m}}{2}{0} -\p{\mathrm{m}}{2}{3}  $ \\
13 & \g4~ & \g5~ & \g8~ & \g8~ & \m2~ & $0-\p{4}{0}{2} -\p{5}{0}{3} $ & $-\p{\mathrm{i}}{0}{2} -\p{\mathrm{i}}{3}{2}  $ & $-\p{\mathrm{m}}{0}{1} -\p{\mathrm{m}}{3}{1}  $ \\
14 & \g5~ & \g4~ & \g8~ & \g8~ & \m2~ & $0+\p{4}{0}{3} -\p{5}{0}{1} $ & $-\p{\mathrm{i}}{0}{1} -\p{\mathrm{i}}{3}{1}  $ & $-\p{\mathrm{m}}{0}{2} -\p{\mathrm{m}}{3}{2}  $ \\
15 & \g2~ & \g1~ & \g2~ & \g8~ & \m1~ & $0-\p{1}{0}{2} +\p{1}{2}{2} $ & $-\p{\mathrm{i}}{0}{2} -\p{\mathrm{i}}{3}{2}  $ & $+\p{\mathrm{m}}{2}{2} -\p{\mathrm{m}}{1}{3}  $ \\
16 & \g2~ & \g1~ & \g3~ & \g8~ & \m1~ & $0-\p{1}{3}{1} -\p{1}{1}{2} $ & $-\p{\mathrm{i}}{2}{1} -\p{\mathrm{i}}{1}{2}  $ & $-\p{\mathrm{m}}{1}{1} -\p{\mathrm{m}}{2}{3}  $ \\
17 & \g2~ & \g2~ & \g3~ & \g8~ & \m1~ & $0-\p{2}{2}{0} -\p{2}{3}{0} +\p{3}{2}{0} $ & $-\p{\mathrm{i}}{2}{0} -\p{\mathrm{i}}{2}{3}  $ & $+\p{\mathrm{m}}{2}{0} +\p{\mathrm{m}}{2}{3}  $ \\
18 & \g2~ & \g4~ & \g1~ & \g8~ & \m1~ & $0-\p{1}{1}{0} +\p{1}{1}{2} -\p{2}{1}{0} $ & $-\p{\mathrm{i}}{1}{0} -\p{\mathrm{i}}{1}{3}  $ & $-\p{\mathrm{m}}{1}{1} +\p{\mathrm{m}}{2}{3}  $ \\
19 & \g2~ & \g4~ & \g1~ & \g8~ & \m2~ & $0-\p{1}{3}{1} -\p{1}{1}{3} $ & $-\p{\mathrm{i}}{2}{1} -\p{\mathrm{i}}{1}{2}  $ & $-\p{\mathrm{m}}{3}{1} +\p{\mathrm{m}}{2}{2}  $ \\
20 & \g3~ & \g4~ & \g1~ & \g8~ & \m1~ & $0-\p{1}{2}{0} +\p{1}{2}{2} -\p{3}{2}{0} $ & $-\p{\mathrm{i}}{2}{0} -\p{\mathrm{i}}{2}{3}  $ & $-\p{\mathrm{m}}{2}{1} -\p{\mathrm{m}}{1}{3}  $ \\
21 & \g3~ & \g5~ & \g1~ & \g8~ & \m2~ & $0-\p{1}{3}{2} -\p{1}{2}{3} $ & $+\p{\mathrm{i}}{2}{1} +\p{\mathrm{i}}{1}{2}  $ & $-\p{\mathrm{m}}{0}{2} -\p{\mathrm{m}}{1}{2}  $ \\
22 & \g4~ & \g4~ & \g5~ & \g8~ & \m2~ & $0-\p{4}{0}{2} -\p{4}{0}{3} +\p{5}{0}{2} $ & $-\p{\mathrm{i}}{0}{2} -\p{\mathrm{i}}{3}{2}  $ & $+\p{\mathrm{m}}{0}{2} +\p{\mathrm{m}}{3}{2}  $ \\
23 & \g5~ & \g1~ & \g4~ & \g8~ & \m2~ & $0+\p{1}{0}{3} +\p{1}{3}{3} +\p{4}{0}{3} -\p{5}{0}{1} $ & $-\p{\mathrm{i}}{0}{1} -\p{\mathrm{i}}{3}{1}  $ & $-\p{\mathrm{m}}{0}{2} -\p{\mathrm{m}}{3}{2}  $ \\
24 & \g3~ & \g1~ & \g1~ & \g2~ & \m1~ & $0+2\p{1}{3}{0} +\p{1}{0}{3} +\p{1}{3}{3} +\p{2}{3}{0} -\p{3}{1}{0} $ & $-\p{\mathrm{i}}{1}{0} -\p{\mathrm{i}}{1}{3}  $ & $-\p{\mathrm{m}}{2}{0} -\p{\mathrm{m}}{2}{3}  $ \\
25 & \g3~ & \g4~ & \g1~ & \g3~ & \m1~ & $0-\p{1}{0}{3} -\p{1}{1}{3} -\p{4}{0}{2} $ & $-\p{\mathrm{i}}{0}{2} -\p{\mathrm{i}}{3}{2}  $ & $+\p{\mathrm{m}}{2}{2} -\p{\mathrm{m}}{1}{3}  $ \\
26 & \g2~ & \g2~ & \g2~ & \g8~ & \m1~ & $0+3\p{2}{1}{0} $ & $+\p{\mathrm{i}}{1}{0} +\p{\mathrm{i}}{1}{3}  $ & $+\p{\mathrm{m}}{1}{0} +\p{\mathrm{m}}{1}{3}  $ \\
27 & \g2~ & \g1~ & \g8~ & \g8~ & \m2~ & $0-\p{1}{1}{0} -\p{1}{1}{3} -\p{2}{1}{0} $ & $-\p{\mathrm{i}}{1}{0} -\p{\mathrm{i}}{1}{3}  $ & $-\p{\mathrm{m}}{1}{1} +\p{\mathrm{m}}{2}{2}  $ \\
\hline\hline
\end{tabular}

\renewcommand{\lasttablecaption}{9.0}

\caption{Gate sequences for tuning the gate set $\op{G} \in \left\{ \g1, \g2, \g3, \g4, \g5 \right\}$. The condition number of the sensitivity matrix $\mat{S}$ is $\cond(\mat{S}) = \lasttablecaption$. The last two sequences are optional and can be used to make the calibration insensitive to offsets \vc{\delta} in the measurement. This is done by subtracting their measurement results from the other sequences with matching measurement operators. After subtraction, the condition number of the difference of the sensitivity matrices is $\cond(\mat{S}_{\mathrm{diff}}) = 83.5$}
\label{tab:diff_seq_all}
\end{table*}

\section{Application to singlet-triplet qubits}
\label{sec:application_to_st_qubits}
In order to study the experimental feasibility and convergence properties of our tuning procedure we now consider a specific qubit model for qubits encoded in two electron spins, so-called \sts qubits \cite{Petta2005}. Previous work has shown that it should be possible to implement quantum gates with fidelities approaching \SI{99.9}{\%} in this system but reaching this level of accuracy experimentally will require tune-up of many parameters \cite{Cerfontaine2019}. GSC was developed with this application in mind. 

\begin{figure}[t]
 	\centering
	\includegraphics[width=0.8\columnwidth]{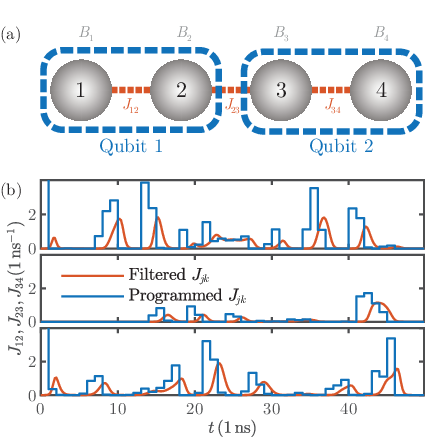}
	\caption{F\textbf{(a)} Diagram of a quadruple quantum dot configuration forming two $S-T_0$ qubits. The local exchange interactions are labeled $J_{12}$ and $J_{34}$, the non-local exchange is labeled $J_{23}$. In addition to the exchange interaction, the spins are subject magnetic fields $B_i$ (due to an effective Overhauser field generated by the host material's nuclear spins and an externally applied magnetic field). \textbf{(b)} Pulse sequence for a \g1 gate with a fidelity of $99.8\%$ (in GaAs) limited by decoherence. This gate is used for the simulations of the gate set calibration protocol. Figure adapted from Ref.\,\citenum{Cerfontaine2019}.}
	\label{fig:qubit}	
\end{figure}

In order to study the calibration of a two-qubit gate, we use the same model as \citet{Cerfontaine2019}, which we now summarize briefly. Specifically, we consider two \sts qubits formed by four linearly adjacent quantum dots as shown in \reffig{fig:qubit}\,(a). Gate electrodes on top of the heterostructure control the chemical potential of the quantum dots and allow control of three exchange interactions $J_{j,j+1}$, $j \in \{1,2,3\}$, between any two neighboring dots via gate voltages $\eps_{j,j+1}$. Each spin also experiences a different magnetic field $B_j$, $j \in \{1,2,3,4\}$. With $\hbar = 1$, the Hamiltonian can then be written as
\begin{align}
\label{eq:hj}
	H = & \sum_{j=1}^3 \frac{J_{j,j+1}}{4}\vecsig ^{(j)} \cdot \vecsig ^{(j+1)} + \frac{1}{2}  \sum_{j=1}^4 B_j \sigma^{(j)},
\end{align}
where $\sigma^{(j)}$ is the Pauli operator acting on the spin in quantum dot $j$. The qubits are encoded in the subspace $\left\{ \ket{\uparrow \downarrow}, \ket{\downarrow \uparrow} \right\}^{\otimes 2}$ where only differences between the magnetic fields, $\db{j, j+1} = B_j-B_{j+1}$, affect the qubit dynamics. Occupation of the states $\ket{\downarrow \downarrow \uparrow \uparrow}$ and $\ket{\uparrow \uparrow \downarrow \downarrow }$ is suppressed to a large extent by choosing $J_{23}/|\db{23}| \ll 1$ \cite{Wardrop2014}. We focus on qubits defined in a GaAs heterostructure, which have been explored more widely as S-T$_0$ qubits and for which we have detailed, validated models at hand. Since experiments in GaAs can achieve \db{j,j+1} from \SI{0.1}{ns^{-1}} to \SI{7}{ns^{-1}} by polarizing the GaAs nuclear spins \cite{Bluhm2010, Nichol2017}, we choose $\db{23} = \SI{7}{ns^{-1}}$ and $\db{12} = -\db{34} = \SI{1}{ns^{-1}}$.

The detunings $\eps_{j,j+1}$ are controlled by using arbitrary waveform generators (AWGs) with a fixed sample rate of typically $\SI{1}{GS/s}$. Typical gates between the four spins consist of $\nseg = 20$ to 50 samples, and are parametrized by three detuning voltages for each sample, $\eps_{j,j+1,l}$, $l=1...\nseg$. From the discrete $\eps_{j,j+1,l}$ we obtain the actual experimental pulse shapes seen by the qubit $\epsilon_{j, j+1}(t)$ by convolution with a typical experimental impulse response \cite{Cerfontaine2019}. Using a phenomenological relation for $J_{j,j+1}(\epsilon_{j,j+1}) = J_0 \exp{(\epsilon_{j,j+1}/\epsilon_0)}$ allows us to obtain $J_{j, j+1}(t)$ using the experimentally measured parameters\cite{Dial2013} $\epsilon_0 = \SI{0.272}{mV}$ and $J_0 = \SI{1}{ns^{-1}}$. Given all \eps values, we can thus calculate the time evolution of this system numerically by approximating the time-dependence of $J_{j, j+1}(t)$ and thus the Hamiltonian as piece-wise constant. Once we have obtained the unitary matrix $U(\eps)$ in this manner, we compare this gate to the desired unitary matrix $U_t$ using the infidelity $\infid(\eps) = 1 - \fid(U(\eps), U_t)$ \cite{Nielsen2002}. In addition, we calculate leakage out of the computational subspace $\leak = 1 - \mathrm{tr}(V_{\mr{c}}^{\dagger}V_{\mr{c}})/4$, where $V_{\mr{c}}$ is the truncation of $U(\eps)$ into the four-dimensional computational subspace. 

We also evaluate the infidelity and leakage due to charge noise affecting \eps and hyperfine noise affecting \db{j,j+1} by calculating the average of $\infid(\eps)$ over 100 computer-generated noise traces. In GaAs, hyperfine noise can be approximated as quasistatic with an experimentally measured standard deviation of $\sigma_{\db{}} = \SI{0.3}{mT}$ \cite{Cerfontaine2016} using dynamic nuclear polarization \cite{Bluhm2010}. For charge noise, we use a quasistatic noise model with $\sigma_{\eps} = \SI{8}{\mu V}$ to which we add white noise with an experimentally measured noise strength of $\SI{4e-20}{V^2/Hz}$ at \SI{1}{MHz} \cite{Dial2013}. 

\sts qubits are typically initialized in a singlet state. They are read out in the singlet-triplet basis by detuning the two adjacent quantum dots, and observing the charge response. For a spin singlet, both electrons will occupy the energetically favourable dot. However, for a triplet state the Pauli exclusion principle applies and the charge state remains unchanged. Thus, singlet and triplet states can be distinguished by their charge response \cite{Barthel2009,Barthel2010} but it is not possible to tell different triplet states apart.

It is possible to apply an adiabatic $\pi/2$ pulse before readout and a $-\pi/2$ pulse after initialization to map $\ket{S}$ to $\ket{\ud}$ and $\ket{T_0}$ to $\ket{\du}$ before readout (and vice-versa for initialization) \cite{Shulman2012}. Thus, we assume that the initial state of the qubits is $\op{\rho}_1 = \ket{\ud\ud}\!\bra{\ud\ud}$. To include realistic measurements, we construct the measurement operators $\op{M}_m$ such that $\ket{\ud}$ states yield a measurement result of 1 and $\ket{\du}$, $\ket{T_+}$, $\ket{T_-}$ yield -1. In the basis $(\ket{\downarrow \downarrow \uparrow \uparrow},
\ket{\uparrow \downarrow \uparrow \downarrow},
\ket{\uparrow \downarrow \downarrow \uparrow},
\ket{\downarrow \uparrow \uparrow \downarrow},
\ket{\downarrow \uparrow \downarrow \uparrow},
\ket{\uparrow \uparrow \downarrow \downarrow })$ the measurement operators $\op{M}_1$ and $\op{M}_2$ for the first and second qubit are then given as
\begin{equation*}
\op{M}_1 =
    \begin{pmatrix}
-1 &   &   \\
  &  \pz \otimes \pid &   \\
  &   &  -1 \\
\end{pmatrix},\,
\op{M}_2 =
    \begin{pmatrix}
-1 &  & \\
&  \pid \otimes \pz & \\
&  &  -1 \\
\end{pmatrix},
\end{equation*}
where omitted matrix elements are zero.

We now simulate two different GSC protocols using sequences 1-15 given in \reftab{tab:reg_seq_cnot_sim} or sequences 1-25 in \reftab{tab:diff_seq_all} with measurements $\m1 = \op{M}_1$ and $\m2 = \op{M}_2$ . While the first case assumes perfect single-qubit gates, which is realistic if they have already been calibrated separately, the second case is truly self-consistent. For both cases, we use the initial state and measurements discussed above, and a previously optimized \SI{50}{ns} long $G_1 = \g1$ gate shown in \reffig{fig:qubit}\,(b) \cite{Cerfontaine2019}. In addition, we use the \SI{20}{ns} long single-qubit gates $G_2 = \g2, G_3 = \g3, G_4 = \g4$ and $G_5 = \g5$ from Ref.\,\citenum{Cerfontaine2019} (without capacitive interqubit coupling and the interqubit exchange $J_{23}$ turned off). The \g1 gate is parametrized by $3 \cdot 50$ free parameters, while each single-qubit gate has $20$ free parameters. For the noise model at hand, the highest predicted gate fidelity for all gates is \SI{99.8}{\%} in the absence of systematic errors. 

Before solving the minimization problem from \refeq{eq:min}, we introduce coherent gate errors by adding random, uncorrelated disturbances to the detunings \eps of the gates subject to calibration. For the single qubit gates we do not perturb \eps on the idling qubit. Thus, we obtain a set of 2500 random starting points with initial infidelities $\infid_\mr{initial}$ of up to $\SI{20}{\%}$ due to systematic errors.

\begin{figure*}[t]
	\centering
	\includegraphics[width=2\columnwidth]{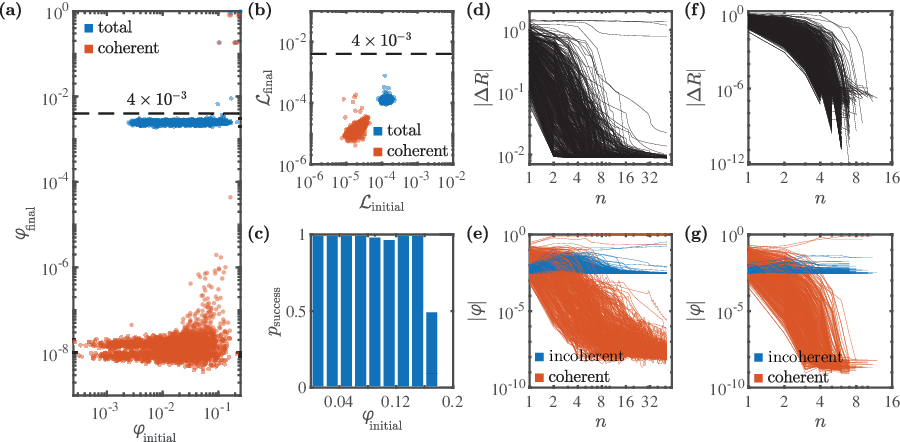}
	\caption{Simulation results for tuning a \g1 gate using all sequences from \reftab{tab:reg_seq_cnot_sim}.
	Leakage \leak and infidelities \infid from coherent errors are shown in red. Blue data also include the effect of decoherence from quasistatic and high-frequency noise. \textbf{(a-b)} Final infidelities and leakages obtained after completing the calibration protocol, as a function of the initial infidelities. The dashed line indicates error and leakage rates of \SI{4e-3}{}. \textbf{(c)} Success rate of the protocol (final infidelity $\ge \SI{99.6}{\%}$) as a function of the initial infidelities. \textbf{(d-e)} Convergence plots for the error syndromes and actual infidelities. \textbf{(f-g)} Convergence plots but for simulation runs without the additional sequences 16 and 17.}
	\label{fig:bootstrap}	
\end{figure*}

\begin{figure}[t]
	\centering
	\includegraphics[width=1\columnwidth]{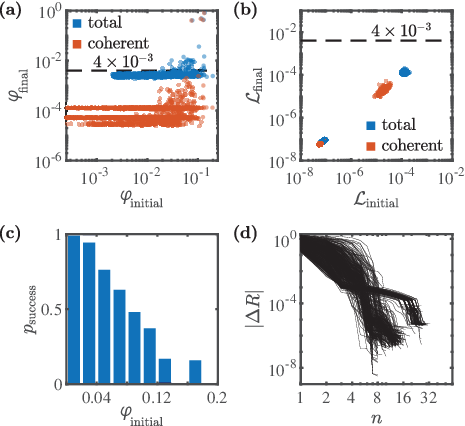}
	\caption{
	Simulation results for tuning a set of four single-qubit gates and a \g1 gate using the sequences from \reftab{tab:diff_seq_all}. \textbf{(a-b)} Final infidelities and leakages obtained after completing the calibration protocol, as a function of the initial infidelities. Each gate in the set is represented by a different data point for each run. \textbf{(c)} Success rate of the protocol (final infidelity of the \g1 gate $\ge \SI{99.6}{\%}$) as a function of the initial infidelities. \textbf{(d)} Convergence plot for the error syndromes.}
	\label{fig:bootstrap_all}	
\end{figure}

For each of these starting points, we iteratively solve the minimization problem from \refeq{eq:min} with respect to \eps using the Levenberg-Marquardt algorithm. The detunings are bounded to experimentally feasible values, $-5.4\eps_0 \le \eps_{i,j+1}(t) \le 2.4\eps_0$. For the single-qubit gates the interqubit exchange is not needed and thus $J_{23}$ is set to 0. In each iteration we use the current values of \eps to construct unitary operators for each gate, which then allows us to calculate the measurement outcomes for each sequence. We do not take quantization noise of the measurements into account since it was shown that with sufficient averaging this is not a fundamental problem for convergence \cite{Cerfontaine2014}. 

For the first case where just the \g1 gate is tuned, we investigate the effect of sequences 16 and 17 which are included to keep decoherence low. To this end, we run two versions of GSC, with and without sequences 16 and 17. For simplicity, we approximate the decoherence channel of each sequence by a depolarizing channel when the two additional sequences are included. In each iteration, we match the infidelity of the depolarizing channel to the current infidelity of the \g1 gate times the number of times the \g1 gate occurs in the respective sequence. Note that a strong anisotropy in the actual decoherence process may require a more refined choice of the decoherence detection sequences to avoid a degraded performance compared to this simple generic model. The infidelity of the \g1 gate due to white charge noise is calculated in a computationally efficient manner by using a Markov approximation \cite{Wardrop2014}, while quasistatic noise is included by integrating over discrete Gaussian distributions \cite{Cerfontaine2019, Grace2013}. When sequences 16 and 17 are not used and in the case where all gates are tuned self-consistently, we limit ourselves to unitary dynamics to make the calculation of derivatives more computationally efficient. Note that the depolarizing channel used during optimization does not lead to systematic measurement offsets $\delta$ for sequences 1 to 15. Thus, difference in the convergence behaviour are only due to the effect of the decoherence sequences 16 and 17.

For each optimization run we characterize the initial and final gate fidelities, \infidi and \infidf, and the infidelity $\infid_k$ in iteration $k$, always including decoherence. These figures are obtained from the noise-afflicted quantum operations calculated in each iteration. An analysis for tuning the \g1 gate with decoherence sequences included is given in \reffig{fig:bootstrap} panels (a)-(e), while we present data obtained without the decoherence sequences in panels (f) and (g). In \reffig{fig:bootstrap_all}, the entire gate set was tuned, and thus we plot the infidelities for all gates. We will now discuss each of the panels in turn. 

In both \reffig{fig:bootstrap} and \reffig{fig:bootstrap_all}, we plot \infidf versus \infidi in panels (a), where red points just contain coherent errors while blue data additionally includes decoherence from noise. The results indicate that most optimization runs successfully remove coherent errors to below \SI{2e-4}{}, consistent with the expectation that some but only small systematic errors arise from decoherence. Much lower coherent errors rates on the order of \SI{1e-8}{} can be reached if only the \g1 gate is tuned. In both cases, incoherent errors cannot be reduced below those of the initial optimized gates. Each of the horizontal clusters seen in \reffig{fig:bootstrap_all} belongs to a different gate being tuned, indicating that coherent errors are removed to a different degree due to the asymmetric construction of the gate sequences with respect to the elements of the gate set. In panels (b) a similar analysis is shown for leakage. Due to a sufficiently low ratio $J_{23}/\db{23}$, leakage is already very low in the beginning, and this property is retained throughout the calibration.

We define a run as successful if the final fidelity of the \g1 gate is higher than $\SI{99.6}{\%}$, close to the optimal value of \SI{99.8}{\%}, and thus obtain the success rates in panels (c). We focus on the \g1 gate because all runs achieved single-qubit gate fidelities above \SI{99.6}{\%}. These result indicate that \g1 gates with initial infidelities up to $\SI{20}{\%}$ can reliably calibrated if single-qubit gates have been calibrated previously. If single-qubit gates are tuned along with the \g1 gate, the success rate drops to about \SI{50}{\%} at \SI{10}{\%} initial gate fidelity. This is probably caused by a greater distance from the local minimum so that convergence can be improved by using algorithms with better global convergence properties. In conclusion, the LMA is less suited for initial tune-up of all gates without further improvements, but will prove useful for compensating smaller changes due to drift.

Next, we investigate the convergence of the error syndromes and infidelities in \reffig{fig:bootstrap} in panel (d) and  (e), respectively. In the beginning, coherent errors are removed quickly, resulting in rapid convergence. Panel (d) in \reffig{fig:bootstrap_all} shows that convergence is similar and takes at most two times more iterations, even though $80$ additional parameters are tuned.

However, we can see from \reffig{fig:bootstrap} (d) and (e) that once coherent errors are comparable to incoherent errors (around iteration 4), convergence slows. This behaviour is not observed in panels (f) and (g) of \reffig{fig:bootstrap}, which show the convergence without the two additional decoherence sequences. We attribute this to the fact that the Levenberg-Marquardt inherits some properties from the Gauss-Newton algorithm (as it interpolates between the Gauss-Newton algorithm and gradient descent). Specifically, 
the Gauss-Newton algorithm can achieve superlinear convergence close to an exact solution, otherwise its convergence is typically not better than linear \cite{Dahmen2008}. If we just include the first 15 sequences, the simulation has an exact solution. In the presence of decoherence there is no exact solution as the decoherence sequences will not be able prepare the ideal pure final state, resulting in the observed slower convergence of the Gauss-Newton algorithm. Thus, additional decoherence sequences should be avoided or removed from the optimization once the convergence slows, unless an optimization algorithm that is less sensitive to this issue is used.

Even with decoherence sequences included, coherent errors can be removed to a level below incoherent errors in around 10 iterations. A previous experiment on single-qubit gate tune-up \cite{Cerfontaine2019ex} measured 2 times fewer sequences and tuned about 3 times fewer parameters in a comparable number of iterations, where each iteration roughly took between 4 and \SI{40}{min}. Tuning a two-qubit gate from an initial fidelity of less than \SI{90}{\%} could take about a factor 6 longer as more measurements are needed to determine gradients using finite differences. While the associated timescale is still experimentally feasible, further speedup would be beneficial. Options for improvement include measuring or approximating the Jacobian in a more efficient manner \cite{Teske2019}, reducing measurement noise or improving the time it takes to upload pulses to the AWG. 

Our approach can also be used in conjunction with derivative-free methods like the Nelder-Mead simplex algorithm \cite{Nelder1964}, which uses a scalar objective function. However, this may turn out to be unfavorable as information on the nature of the errors is lost. While using a derivative-free algorithm reduces the number of required measurements, it also leads to slower convergence. For example, Ref.\,\citenum{Kelly2014} needed 80 iterations to improve a gate fidelity from \SI{98.4}{\%} to \SI{99.3}{\%}. Determining which approach is optimal and faster depends on the details of the experiment. Aspects to consider include the noise levels present in the system, the number of gate parameters to be tuned and the initial gate fidelity before tuning. 

\section{Conclusion}
\label{sec:conclusion}
In this work, we have introduced GSC, a gate set calibration protocol which allows for straightforward experimental calibration of different gate sets on one or several qubits. We have presented explicit sequences for a two-qubit gate set but it is straightforward to automatically generate sequences optimally suited for a range of gate sets, experimental requirements and specific errors models.

Furthermore, it is possible to construct gate sequences robust to some forms of state preparation and measurement errors and decoherence. This is advantageous if some errors are more likely to occur than others. We have given explicit gate sequences which are first-order sensitive to only one coherent single-qubit readout error. From our simulations, we also expect that our method will be able to tune up an entire two-qubit gate set for \sts spin-qubits. Even if 230 or more free parameters need be adjusted, fast convergence is possible. As discussed, the moderate resource requirements of GSC can also be harnessed to calibrate classical and non-classical cross-talk in multi-qubit architectures. 

To consider larger qubit systems and longer gate sequences, the current search for suitable sequences could also be replaced by a more structured approach. This can include an explicit optimization of the information gained by each measurement, instead of simply constructing a well-conditioned matrix, which will enable longer gate sequences to amplify certain errors by multiple application of the same gate. Furthermore, it will be useful to investigate methods for making the gates sequences more robust to decoherence and SPAM errors, and extracting precise errors bars on the error syndromes.

While GSC exhibits a nearly ideal performance in terms of convergence, it might be necessary to explore other approaches if the qubit system does not allow for initial infidelities below $10 - 20\,\%$. These could include a coarse pre-calibration by reconstructing state trajectories and optimizing their agreement with the ideal case, or abandoning the idea of starting from simulated seeds and doing a global search on the experiment instead.

\begin{acknowledgments}
This work was supported by the European Research Council (ERC) under the European Union’s Horizon 2020 research and innovation program (grant agreement No. 679342) and the Impulse and Networking Fund of the Helmholtz Association. P.C. acknowledges support by Deutsche Telekom Stiftung. We appreciate valuable discussions with Manuel Rispler.

Reprinted with permission from Pascal Cerfontaine, René Otten, and Hendrik Bluhm, Physical Review Applied 13, 044071 (2020). Copyright 2020 by the American Physical Society.
\end{acknowledgments}

\clearpage
\newpage
\onecolumngrid
\appendix
\section*{Additional gate sequences}
In this appendix, we provide the additional gate sequences referenced in the main text. The first two tables, \reftab{tab:diff_seq_cnot} and \reftab{tab:diff_seq_cnot_2}, belong together. By subtracting the measurement results obtained by using the sequences from the second table from those obtained from the first table, common measurement offsets can be cancelled. Each sequence pair prepares the same final state. The same procedure should be used with the sequences from the last two tables, \reftab{tab:two_added_seq_cnot} and \reftab{tab:two_added_seq_cnot_2}, which only differ by two additional gates in each sequence.

\begin{table*}[b!]

\centering
\renewcommand\arraystretch{1.3}
\begin{tabular}{r *{9}{l}}
\hline\hline
s & $\gamma_{s,1}$ & $\gamma_{s,2}$ & $\gamma_{s,3}$ & $\gamma_{s,4}$ & $\gamma_{s,5}$ & $\op{M}_s$ & $R_r(\frac{1}{2} \vc{p})$ for Gates & Initialization & Measurement \\
\hline
1 & \g1~ & \g2~ & \g8~ & \g8~ & \g8~ & \m1~ & $0-\p{1}{1}{1} +\p{1}{2}{2} $ & $-\p{\mathrm{i}}{1}{1} +\p{\mathrm{i}}{2}{2}  $ & $-\p{\mathrm{m}}{1}{0} -\p{\mathrm{m}}{1}{3}  $ \\
2 & \g1~ & \g3~ & \g8~ & \g8~ & \g8~ & \m1~ & $0-\p{1}{2}{1} -\p{1}{1}{2} $ & $-\p{\mathrm{i}}{2}{1} -\p{\mathrm{i}}{1}{2}  $ & $-\p{\mathrm{m}}{2}{0} -\p{\mathrm{m}}{2}{3}  $ \\
3 & \g1~ & \g4~ & \g8~ & \g8~ & \g8~ & \m2~ & $0-\p{1}{0}{1} -\p{1}{3}{1} $ & $-\p{\mathrm{i}}{0}{1} -\p{\mathrm{i}}{3}{1}  $ & $-\p{\mathrm{m}}{0}{1} -\p{\mathrm{m}}{3}{1}  $ \\
4 & \g1~ & \g5~ & \g8~ & \g8~ & \g8~ & \m2~ & $0-\p{1}{0}{2} -\p{1}{3}{2} $ & $-\p{\mathrm{i}}{0}{2} -\p{\mathrm{i}}{3}{2}  $ & $-\p{\mathrm{m}}{0}{2} -\p{\mathrm{m}}{3}{2}  $ \\
5 & \g2~ & \g1~ & \g8~ & \g8~ & \g8~ & \m1~ & $0-\p{1}{1}{0} -\p{1}{1}{3} $ & $-\p{\mathrm{i}}{1}{0} -\p{\mathrm{i}}{1}{3}  $ & $-\p{\mathrm{m}}{1}{1} +\p{\mathrm{m}}{2}{2}  $ \\
6 & \g3~ & \g1~ & \g8~ & \g8~ & \g8~ & \m1~ & $0-\p{1}{2}{0} -\p{1}{2}{3} $ & $-\p{\mathrm{i}}{2}{0} -\p{\mathrm{i}}{2}{3}  $ & $-\p{\mathrm{m}}{2}{1} -\p{\mathrm{m}}{1}{2}  $ \\
7 & \g2~ & \g1~ & \g2~ & \g8~ & \g8~ & \m1~ & $0-\p{1}{0}{2} +\p{1}{2}{2} $ & $-\p{\mathrm{i}}{0}{2} -\p{\mathrm{i}}{3}{2}  $ & $+\p{\mathrm{m}}{2}{2} -\p{\mathrm{m}}{1}{3}  $ \\
8 & \g2~ & \g1~ & \g3~ & \g8~ & \g8~ & \m1~ & $0-\p{1}{3}{1} -\p{1}{1}{2} $ & $-\p{\mathrm{i}}{2}{1} -\p{\mathrm{i}}{1}{2}  $ & $-\p{\mathrm{m}}{1}{1} -\p{\mathrm{m}}{2}{3}  $ \\
9 & \g2~ & \g4~ & \g1~ & \g8~ & \g8~ & \m1~ & $0-\p{1}{1}{0} +\p{1}{1}{2} $ & $-\p{\mathrm{i}}{1}{0} -\p{\mathrm{i}}{1}{3}  $ & $-\p{\mathrm{m}}{1}{1} +\p{\mathrm{m}}{2}{3}  $ \\
10 & \g2~ & \g4~ & \g1~ & \g8~ & \g8~ & \m2~ & $0-\p{1}{3}{1} -\p{1}{1}{3} $ & $-\p{\mathrm{i}}{2}{1} -\p{\mathrm{i}}{1}{2}  $ & $-\p{\mathrm{m}}{3}{1} +\p{\mathrm{m}}{2}{2}  $ \\
11 & \g3~ & \g4~ & \g1~ & \g8~ & \g8~ & \m1~ & $0-\p{1}{2}{0} +\p{1}{2}{2} $ & $-\p{\mathrm{i}}{2}{0} -\p{\mathrm{i}}{2}{3}  $ & $-\p{\mathrm{m}}{2}{1} -\p{\mathrm{m}}{1}{3}  $ \\
12 & \g3~ & \g5~ & \g1~ & \g8~ & \g8~ & \m2~ & $0-\p{1}{3}{2} -\p{1}{2}{3} $ & $+\p{\mathrm{i}}{2}{1} +\p{\mathrm{i}}{1}{2}  $ & $-\p{\mathrm{m}}{0}{2} -\p{\mathrm{m}}{1}{2}  $ \\
13 & \g4~ & \g1~ & \g5~ & \g8~ & \g8~ & \m2~ & $0-\p{1}{0}{3} -\p{1}{3}{3} $ & $-\p{\mathrm{i}}{0}{2} -\p{\mathrm{i}}{3}{2}  $ & $-\p{\mathrm{m}}{0}{1} -\p{\mathrm{m}}{3}{1}  $ \\
14 & \g2~ & \g1~ & \g1~ & \g3~ & \g8~ & \m1~ & $0-2\p{1}{3}{0} -\p{1}{0}{3} -\p{1}{3}{3} $ & $-\p{\mathrm{i}}{2}{0} -\p{\mathrm{i}}{2}{3}  $ & $-\p{\mathrm{m}}{1}{0} -\p{\mathrm{m}}{1}{3}  $ \\
15 & \g2~ & \g5~ & \g1~ & \g4~ & \g8~ & \m2~ & $0-\p{1}{1}{2} +\p{1}{3}{3} $ & $-\p{\mathrm{i}}{2}{1} -\p{\mathrm{i}}{1}{2}  $ & $-\p{\mathrm{m}}{0}{2} +\p{\mathrm{m}}{2}{2}  $ \\
\hline\hline
\end{tabular}

\renewcommand{\lasttablecaption}{6.8}

\caption{Gate sequences for tuning a \g1 gate. The condition number of the sensitivity matrix $\mat{S}$ is $\cond(\mat{S}) = \lasttablecaption$. These sequences can be used to make the calibration insensitive to offsets in the measurement by subtracting the measurement results of the sequences in \reftab{tab:diff_seq_cnot_2}.}
\label{tab:diff_seq_cnot}
\end{table*}

\begin{table*}[b!]

\centering
\renewcommand\arraystretch{1.3}
\begin{tabular}{r *{9}{l}}
\hline\hline
s & $\gamma_{s,1}$ & $\gamma_{s,2}$ & $\gamma_{s,3}$ & $\gamma_{s,4}$ & $\gamma_{s,5}$ & $\op{M}_s$ & $R_r(\frac{1}{2} \vc{p})$ for Gates & Initialization & Measurement \\
\hline
1 & \g2~ & \g3~ & \g8~ & \g8~ & \g8~ & \m1~ & $0$ & $-\p{\mathrm{i}}{2}{0} -\p{\mathrm{i}}{2}{3}  $ & $-\p{\mathrm{m}}{1}{0} -\p{\mathrm{m}}{1}{3}  $ \\
2 & \g3~ & \g2~ & \g8~ & \g8~ & \g8~ & \m1~ & $0$ & $-\p{\mathrm{i}}{1}{0} -\p{\mathrm{i}}{1}{3}  $ & $-\p{\mathrm{m}}{2}{0} -\p{\mathrm{m}}{2}{3}  $ \\
3 & \g4~ & \g5~ & \g8~ & \g8~ & \g8~ & \m2~ & $0$ & $-\p{\mathrm{i}}{0}{2} -\p{\mathrm{i}}{3}{2}  $ & $-\p{\mathrm{m}}{0}{1} -\p{\mathrm{m}}{3}{1}  $ \\
4 & \g5~ & \g4~ & \g8~ & \g8~ & \g8~ & \m2~ & $0$ & $-\p{\mathrm{i}}{0}{1} -\p{\mathrm{i}}{3}{1}  $ & $-\p{\mathrm{m}}{0}{2} -\p{\mathrm{m}}{3}{2}  $ \\
5 & \g1~ & \g1~ & \g2~ & \g1~ & \g8~ & \m1~ & $0-2\p{1}{1}{0} -\p{1}{1}{1} +\p{1}{2}{2} -2\p{1}{1}{3} $ & $-\p{\mathrm{i}}{1}{0} -\p{\mathrm{i}}{1}{3}  $ & $-\p{\mathrm{m}}{1}{1} +\p{\mathrm{m}}{2}{2}  $ \\
6 & \g1~ & \g1~ & \g3~ & \g1~ & \g8~ & \m1~ & $0-2\p{1}{2}{0} -\p{1}{2}{1} -\p{1}{1}{2} -2\p{1}{2}{3} $ & $-\p{\mathrm{i}}{2}{0} -\p{\mathrm{i}}{2}{3}  $ & $-\p{\mathrm{m}}{2}{1} -\p{\mathrm{m}}{1}{2}  $ \\
7 & \g2~ & \g1~ & \g2~ & \g1~ & \g8~ & \m1~ & $0-\p{1}{0}{2} +2\p{1}{2}{2} -\p{1}{1}{3} $ & $-\p{\mathrm{i}}{0}{2} -\p{\mathrm{i}}{3}{2}  $ & $+\p{\mathrm{m}}{2}{2} -\p{\mathrm{m}}{1}{3}  $ \\
8 & \g2~ & \g1~ & \g1~ & \g1~ & \g3~ & \m1~ & $0-3\p{1}{3}{1} -2\p{1}{1}{2} -\p{1}{2}{3} $ & $-\p{\mathrm{i}}{2}{1} -\p{\mathrm{i}}{1}{2}  $ & $-\p{\mathrm{m}}{1}{1} -\p{\mathrm{m}}{2}{3}  $ \\
9 & \g2~ & \g1~ & \g4~ & \g8~ & \g8~ & \m1~ & $0-\p{1}{1}{0} -\p{1}{1}{3} $ & $-\p{\mathrm{i}}{1}{0} -\p{\mathrm{i}}{1}{3}  $ & $-\p{\mathrm{m}}{1}{1} +\p{\mathrm{m}}{2}{3}  $ \\
10 & \g2~ & \g1~ & \g4~ & \g8~ & \g8~ & \m2~ & $0-\p{1}{3}{1} -\p{1}{1}{2} $ & $-\p{\mathrm{i}}{2}{1} -\p{\mathrm{i}}{1}{2}  $ & $-\p{\mathrm{m}}{3}{1} +\p{\mathrm{m}}{2}{2}  $ \\
11 & \g3~ & \g1~ & \g2~ & \g8~ & \g8~ & \m1~ & $0+\p{1}{3}{1} +\p{1}{2}{2} $ & $-\p{\mathrm{i}}{1}{1} +\p{\mathrm{i}}{2}{2}  $ & $-\p{\mathrm{m}}{2}{1} -\p{\mathrm{m}}{1}{3}  $ \\
12 & \g1~ & \g3~ & \g5~ & \g8~ & \g8~ & \m2~ & $0-\p{1}{0}{2} -\p{1}{3}{2} $ & $-\p{\mathrm{i}}{0}{2} -\p{\mathrm{i}}{3}{2}  $ & $-\p{\mathrm{m}}{0}{2} -\p{\mathrm{m}}{1}{2}  $ \\
13 & \g1~ & \g1~ & \g4~ & \g8~ & \g8~ & \m2~ & $0-2\p{1}{0}{1} -2\p{1}{3}{1} $ & $-\p{\mathrm{i}}{0}{1} -\p{\mathrm{i}}{3}{1}  $ & $-\p{\mathrm{m}}{0}{1} -\p{\mathrm{m}}{3}{1}  $ \\
14 & \g1~ & \g1~ & \g1~ & \g2~ & \g8~ & \m1~ & $0-\p{1}{1}{0} -2\p{1}{1}{1} +2\p{1}{2}{2} -\p{1}{1}{3} $ & $-\p{\mathrm{i}}{1}{1} +\p{\mathrm{i}}{2}{2}  $ & $-\p{\mathrm{m}}{1}{0} -\p{\mathrm{m}}{1}{3}  $ \\
15 & \g1~ & \g1~ & \g2~ & \g5~ & \g8~ & \m2~ & $0-2\p{1}{0}{2} -2\p{1}{3}{2} $ & $-\p{\mathrm{i}}{0}{2} -\p{\mathrm{i}}{3}{2}  $ & $-\p{\mathrm{m}}{0}{2} +\p{\mathrm{m}}{2}{2}  $ \\
\hline\hline
\end{tabular}

\renewcommand{\lasttablecaption}{17.9}

\caption{Gate sequences for tuning a \g1 gate. These sequences can be used to make the calibration insensitive to offsets in the measurement by subtracting their measurement results from those of the sequences in \reftab{tab:diff_seq_cnot}. In the absence of coherent errors, each sequence prepares the same final state as the corresponding sequence in \reftab{tab:diff_seq_cnot}. After subtraction, the condition number of the difference of the sensitivity matrices is $\cond(\mat{S}_{\mathrm{diff}}) = \lasttablecaption$.}
\label{tab:diff_seq_cnot_2}
\end{table*}

\begin{table*}[b!]

\centering
\renewcommand\arraystretch{1.3}
\begin{tabular}{r *{10}{l}}
\hline\hline
s & $\gamma_{s,1}$ & $\gamma_{s,2}$ & $\gamma_{s,3}$ & $\gamma_{s,4}$ & $\gamma_{s,5}$ & $\gamma_{s,6}$ & $\op{M}_s$ & $R_r(\frac{1}{2} \vc{p})$ for Gates & Initialization & Measurement \\
\hline
1 & \g1~ & \g2~ & \g8~ & \g8~ & \g8~ & \g8~ & \m1~ & $0-\p{1}{1}{1} +\p{1}{2}{2} $ & $-\p{\mathrm{i}}{1}{1} +\p{\mathrm{i}}{2}{2}  $ & $-\p{\mathrm{m}}{1}{0} -\p{\mathrm{m}}{1}{3}  $ \\
2 & \g1~ & \g3~ & \g8~ & \g8~ & \g8~ & \g8~ & \m1~ & $0-\p{1}{2}{1} -\p{1}{1}{2} $ & $-\p{\mathrm{i}}{2}{1} -\p{\mathrm{i}}{1}{2}  $ & $-\p{\mathrm{m}}{2}{0} -\p{\mathrm{m}}{2}{3}  $ \\
3 & \g1~ & \g4~ & \g8~ & \g8~ & \g8~ & \g8~ & \m2~ & $0-\p{1}{0}{1} -\p{1}{3}{1} $ & $-\p{\mathrm{i}}{0}{1} -\p{\mathrm{i}}{3}{1}  $ & $-\p{\mathrm{m}}{0}{1} -\p{\mathrm{m}}{3}{1}  $ \\
4 & \g1~ & \g5~ & \g8~ & \g8~ & \g8~ & \g8~ & \m2~ & $0-\p{1}{0}{2} -\p{1}{3}{2} $ & $-\p{\mathrm{i}}{0}{2} -\p{\mathrm{i}}{3}{2}  $ & $-\p{\mathrm{m}}{0}{2} -\p{\mathrm{m}}{3}{2}  $ \\
5 & \g2~ & \g1~ & \g8~ & \g8~ & \g8~ & \g8~ & \m1~ & $0-\p{1}{1}{0} -\p{1}{1}{3} $ & $-\p{\mathrm{i}}{1}{0} -\p{\mathrm{i}}{1}{3}  $ & $-\p{\mathrm{m}}{1}{1} +\p{\mathrm{m}}{2}{2}  $ \\
6 & \g3~ & \g1~ & \g8~ & \g8~ & \g8~ & \g8~ & \m1~ & $0-\p{1}{2}{0} -\p{1}{2}{3} $ & $-\p{\mathrm{i}}{2}{0} -\p{\mathrm{i}}{2}{3}  $ & $-\p{\mathrm{m}}{2}{1} -\p{\mathrm{m}}{1}{2}  $ \\
7 & \g2~ & \g1~ & \g3~ & \g8~ & \g8~ & \g8~ & \m1~ & $0-\p{1}{3}{1} -\p{1}{1}{2} $ & $-\p{\mathrm{i}}{2}{1} -\p{\mathrm{i}}{1}{2}  $ & $-\p{\mathrm{m}}{1}{1} -\p{\mathrm{m}}{2}{3}  $ \\
8 & \g2~ & \g4~ & \g1~ & \g8~ & \g8~ & \g8~ & \m2~ & $0-\p{1}{3}{1} -\p{1}{1}{3} $ & $-\p{\mathrm{i}}{2}{1} -\p{\mathrm{i}}{1}{2}  $ & $-\p{\mathrm{m}}{3}{1} +\p{\mathrm{m}}{2}{2}  $ \\
9 & \g2~ & \g5~ & \g1~ & \g8~ & \g8~ & \g8~ & \m1~ & $0-\p{1}{1}{0} -\p{1}{1}{1} $ & $-\p{\mathrm{i}}{1}{0} -\p{\mathrm{i}}{1}{3}  $ & $-\p{\mathrm{m}}{1}{0} -\p{\mathrm{m}}{1}{1}  $ \\
10 & \g2~ & \g5~ & \g1~ & \g8~ & \g8~ & \g8~ & \m2~ & $0-\p{1}{3}{2} -\p{1}{1}{3} $ & $+\p{\mathrm{i}}{1}{1} -\p{\mathrm{i}}{2}{2}  $ & $-\p{\mathrm{m}}{0}{2} +\p{\mathrm{m}}{2}{2}  $ \\
11 & \g3~ & \g5~ & \g1~ & \g8~ & \g8~ & \g8~ & \m1~ & $0-\p{1}{2}{0} -\p{1}{2}{1} $ & $-\p{\mathrm{i}}{2}{0} -\p{\mathrm{i}}{2}{3}  $ & $-\p{\mathrm{m}}{2}{0} -\p{\mathrm{m}}{2}{1}  $ \\
12 & \g3~ & \g5~ & \g1~ & \g8~ & \g8~ & \g8~ & \m2~ & $0-\p{1}{3}{2} -\p{1}{2}{3} $ & $+\p{\mathrm{i}}{2}{1} +\p{\mathrm{i}}{1}{2}  $ & $-\p{\mathrm{m}}{0}{2} -\p{\mathrm{m}}{1}{2}  $ \\
13 & \g4~ & \g1~ & \g5~ & \g8~ & \g8~ & \g8~ & \m2~ & $0-\p{1}{0}{3} -\p{1}{3}{3} $ & $-\p{\mathrm{i}}{0}{2} -\p{\mathrm{i}}{3}{2}  $ & $-\p{\mathrm{m}}{0}{1} -\p{\mathrm{m}}{3}{1}  $ \\
14 & \g2~ & \g1~ & \g1~ & \g3~ & \g8~ & \g8~ & \m1~ & $0-2\p{1}{3}{0} -\p{1}{0}{3} -\p{1}{3}{3} $ & $-\p{\mathrm{i}}{2}{0} -\p{\mathrm{i}}{2}{3}  $ & $-\p{\mathrm{m}}{1}{0} -\p{\mathrm{m}}{1}{3}  $ \\
15 & \g3~ & \g4~ & \g1~ & \g3~ & \g8~ & \g8~ & \m1~ & $0-\p{1}{0}{3} -\p{1}{1}{3} $ & $-\p{\mathrm{i}}{0}{2} -\p{\mathrm{i}}{3}{2}  $ & $+\p{\mathrm{m}}{2}{2} -\p{\mathrm{m}}{1}{3}  $ \\
\hline\hline
\end{tabular}

\renewcommand{\lasttablecaption}{11.4}

\caption{Gate sequences for tuning a \g1 gate. The condition number of the sensitivity matrix $\mat{S}$ is $\cond(\mat{S}) = \lasttablecaption$. These sequences can be used to make the calibration insensitive to offsets in the measurement by subtracting the measurement results of the sequences in \reftab{tab:two_added_seq_cnot_2}.}
\label{tab:two_added_seq_cnot}
\end{table*}

\begin{table*}[b!]

\centering
\renewcommand\arraystretch{1.3}
\begin{tabular}{r *{10}{l}}
\hline\hline
s & $\gamma_{s,1}$ & $\gamma_{s,2}$ & $\gamma_{s,3}$ & $\gamma_{s,4}$ & $\gamma_{s,5}$ & $\gamma_{s,6}$ & $\op{M}_s$ & $R_r(\frac{1}{2} \vc{p})$ for Gates & Initialization & Measurement \\
\hline
1 & \g1~ & \g2~ & \g8~ & \g8~ & \g3~ & \g5~ & \m1~ & $0-\p{1}{2}{1} -\p{1}{1}{2} $ & $-\p{\mathrm{i}}{1}{1} +\p{\mathrm{i}}{2}{2}  $ & $-\p{\mathrm{m}}{1}{0} -\p{\mathrm{m}}{1}{3}  $ \\
2 & \g1~ & \g3~ & \g8~ & \g8~ & \g1~ & \g8~ & \m1~ & $0-\p{1}{2}{0} -\p{1}{2}{1} -\p{1}{1}{2} -\p{1}{2}{3} $ & $-\p{\mathrm{i}}{2}{1} -\p{\mathrm{i}}{1}{2}  $ & $-\p{\mathrm{m}}{2}{0} -\p{\mathrm{m}}{2}{3}  $ \\
3 & \g1~ & \g4~ & \g8~ & \g8~ & \g5~ & \g4~ & \m2~ & $0-\p{1}{0}{2} -\p{1}{3}{2} $ & $-\p{\mathrm{i}}{0}{1} -\p{\mathrm{i}}{3}{1}  $ & $-\p{\mathrm{m}}{0}{1} -\p{\mathrm{m}}{3}{1}  $ \\
4 & \g1~ & \g5~ & \g8~ & \g8~ & \g1~ & \g3~ & \m2~ & $0-2\p{1}{0}{2} -2\p{1}{3}{2} $ & $-\p{\mathrm{i}}{0}{2} -\p{\mathrm{i}}{3}{2}  $ & $-\p{\mathrm{m}}{0}{2} -\p{\mathrm{m}}{3}{2}  $ \\
5 & \g2~ & \g1~ & \g8~ & \g8~ & \g2~ & \g3~ & \m1~ & $0-\p{1}{0}{2} +\p{1}{2}{2} $ & $-\p{\mathrm{i}}{1}{0} -\p{\mathrm{i}}{1}{3}  $ & $-\p{\mathrm{m}}{1}{1} +\p{\mathrm{m}}{2}{2}  $ \\
6 & \g3~ & \g1~ & \g8~ & \g8~ & \g3~ & \g1~ & \m1~ & $0-\p{1}{0}{2} -\p{1}{1}{2} $ & $-\p{\mathrm{i}}{2}{0} -\p{\mathrm{i}}{2}{3}  $ & $-\p{\mathrm{m}}{2}{1} -\p{\mathrm{m}}{1}{2}  $ \\
7 & \g2~ & \g1~ & \g3~ & \g8~ & \g2~ & \g3~ & \m1~ & $0-\p{1}{0}{2} +\p{1}{2}{2} $ & $-\p{\mathrm{i}}{2}{1} -\p{\mathrm{i}}{1}{2}  $ & $-\p{\mathrm{m}}{1}{1} -\p{\mathrm{m}}{2}{3}  $ \\
8 & \g2~ & \g4~ & \g1~ & \g8~ & \g5~ & \g3~ & \m2~ & $0-\p{1}{0}{3} +\p{1}{2}{3} $ & $-\p{\mathrm{i}}{2}{1} -\p{\mathrm{i}}{1}{2}  $ & $-\p{\mathrm{m}}{3}{1} +\p{\mathrm{m}}{2}{2}  $ \\
9 & \g2~ & \g5~ & \g1~ & \g8~ & \g3~ & \g5~ & \m1~ & $0-\p{1}{3}{0} -\p{1}{3}{1} $ & $-\p{\mathrm{i}}{1}{0} -\p{\mathrm{i}}{1}{3}  $ & $-\p{\mathrm{m}}{1}{0} -\p{\mathrm{m}}{1}{1}  $ \\
10 & \g2~ & \g5~ & \g1~ & \g8~ & \g1~ & \g3~ & \m2~ & $0-\p{1}{0}{2} +\p{1}{2}{2} -\p{1}{3}{2} -\p{1}{1}{3} $ & $+\p{\mathrm{i}}{1}{1} -\p{\mathrm{i}}{2}{2}  $ & $-\p{\mathrm{m}}{0}{2} +\p{\mathrm{m}}{2}{2}  $ \\
11 & \g3~ & \g5~ & \g1~ & \g8~ & \g1~ & \g4~ & \m1~ & $0-2\p{1}{2}{0} -2\p{1}{2}{1} $ & $-\p{\mathrm{i}}{2}{0} -\p{\mathrm{i}}{2}{3}  $ & $-\p{\mathrm{m}}{2}{0} -\p{\mathrm{m}}{2}{1}  $ \\
12 & \g3~ & \g5~ & \g1~ & \g8~ & \g4~ & \g8~ & \m2~ & $0-\p{1}{2}{2} +\p{1}{3}{3} $ & $+\p{\mathrm{i}}{2}{1} +\p{\mathrm{i}}{1}{2}  $ & $-\p{\mathrm{m}}{0}{2} -\p{\mathrm{m}}{1}{2}  $ \\
13 & \g4~ & \g1~ & \g5~ & \g8~ & \g5~ & \g1~ & \m2~ & $0+\p{1}{0}{1} +\p{1}{3}{1} $ & $-\p{\mathrm{i}}{0}{2} -\p{\mathrm{i}}{3}{2}  $ & $-\p{\mathrm{m}}{0}{1} -\p{\mathrm{m}}{3}{1}  $ \\
14 & \g2~ & \g1~ & \g1~ & \g3~ & \g3~ & \g5~ & \m1~ & $0+\p{1}{1}{0} +\p{1}{1}{1} -\p{1}{2}{2} +\p{1}{1}{3} $ & $-\p{\mathrm{i}}{2}{0} -\p{\mathrm{i}}{2}{3}  $ & $-\p{\mathrm{m}}{1}{0} -\p{\mathrm{m}}{1}{3}  $ \\
15 & \g3~ & \g4~ & \g1~ & \g3~ & \g3~ & \g8~ & \m1~ & $0+\p{1}{2}{0} -\p{1}{2}{2} $ & $-\p{\mathrm{i}}{0}{2} -\p{\mathrm{i}}{3}{2}  $ & $+\p{\mathrm{m}}{2}{2} -\p{\mathrm{m}}{1}{3}  $ \\
\hline\hline
\end{tabular}

\renewcommand{\lasttablecaption}{23.0}

\caption{Gate sequences for tuning a \g1 gate. These sequences can be used to make the calibration insensitive to offsets in the measurement by subtracting their measurement results from those of the sequences in \reftab{tab:two_added_seq_cnot}. After subtraction, the condition number of the difference of the sensitivity matrices is $\cond(\mat{S}_{\mathrm{diff}}) = \lasttablecaption$. These sequences are the same as in \reftab{tab:two_added_seq_cnot} except for two additional gates, $\gamma_{s,5}$ and $\gamma_{s,6}$}
\label{tab:two_added_seq_cnot_2}
\end{table*}

\end{document}